\pdfoutput=1

\documentclass[ amsmath, amssymb, aps, prd, twocolumn, lengthcheck, sort&compress, showpacs, superscriptaddress, nofootinbib ]{revtex4-1}

\usepackage{amsmath}
\usepackage{amssymb}
\usepackage{graphicx}
\usepackage{dsfont}
\usepackage{dcolumn}
\usepackage{units}
\usepackage{bm}

\usepackage[usenames]{color}

\usepackage[colorlinks=true,linkcolor=blue,citecolor=blue,urlcolor=blue]{hyperref}

      \newcommand{\conjg}[1]{\ensuremath{\hspace{1pt}\overline{\hspace{-1pt}#1\hspace{-1pt}}}\hspace{1pt}}
      \newcommand{\vect}[1]{\bm{#1}}

      \newcommand{\grad}{\vect{\nabla}}

         \newcommand{\be}{\begin{equation}}
         \newcommand{\ee}{\end{equation}}
         \newcommand{\one}{\ensuremath{\mathbf{1}}}

\def\Slash#1{\setbox0=\hbox{$#1$} % set a box for #1
\dimen0=\wd0 % and get its size
\setbox1=\hbox{/} \dimen1=\wd1 % get size of /
\ifdim\dimen0>\dimen1 % #1 is bigger
\rlap{\hbox to \dimen0{\hfil/\hfil}} % so center / in box
#1 % and print #1
\else % / is bigger
\rlap{\hbox to \dimen1{\hfil$#1$\hfil}} % so center #1
/ % and print /
\fi}

\def\longlonglongrightarrow{
\relbar\joinrel\relbar\joinrel\relbar\joinrel\relbar\joinrel\relbar\joinrel\relbar\joinrel\rightarrow}

\begin{document}

         \title{Nucleon to Delta electromagnetic transition in the Dyson-Schwinger approach}

         \author{G.~Eichmann}
         \affiliation{Institut f\"{u}r Theoretische Physik I, Justus-Liebig-Universit\"at Giessen, D-35392 Giessen, Germany  }
         \email{gernot.eichmann@theo.physik.uni-giessen.de}

         \author{D.~Nicmorus}
         \affiliation{GSI Helmholtzzentrum f\"{u}r Schwerionenforschung, Planckstrasse 1, D-64291 Darmstadt, Germany}
	 \email{nicmorus@th.physik.uni-frankfurt.de}

         \date{\today}

         \begin{abstract}

              We study the $N\Delta\gamma$ transition in the Dyson-Schwinger approach.
              The nucleon and $\Delta$ baryons are treated as quark-diquark bound states, where
              the ingredients of the electromagnetic transition current are computed self-consistently from the underlying dynamics in QCD.
              Although our approach does not include pion-cloud effects, we find that the electric and Coulomb quadrupole form-factor ratios $R_{EM}$ and $R_{SM}$
              show good agreement with experimental data. This implies that the deformation from a spherical charge distribution inside both baryons can be traced back to the appearance of $p$ waves in the
              nucleon and $\Delta$ bound-state amplitudes which are a consequence of Poincar\'e covariance.
              On the other hand, the dominant transition amplitude, i.e. the magnetic dipole transition form factor, underestimates the data by $\sim 25\%$ in the static limit
              whereas agreement is achieved at larger momentum transfer, which is consistent with missing pion-cloud contributions.
              We furthermore find that the static properties of the form factors are not very sensitive to a variation of the current-quark mass.

         \end{abstract}


         \maketitle

%%%%%%%%%%%%%%%%%%%%%%%%%%%%%%%%%%%%%%%%%%%%%%%%%%%%%%%%%%%%%%%%%%%%%%%%%%%%%%%%%%%%%%%%%%%%%%%%%%%%%%%%%%%%%%%%%%%%%%%%%%%%

\section{Introduction}

        Investigating the structure of the nucleon represents a challenging task
        in contemporary hadron physics from both experimental and theoretical perspectives.
        Unambiguous testimony of the nucleon's complex non-pointlike structure is given by
        measurements of the electromagnetic transition into its lowest-lying resonance, the $\Delta$ baryon.
        The properties of the electromagnetic \mbox{spin-$\nicefrac{3}{2}$}--spin-$\nicefrac{1}{2}$ transition are experimentally accessible via
        pion photo- and electroproduction off nucleon targets.
        Precision measurements of the $N\to\Delta\gamma$ transition over a wide range of $Q^2$ have become available only in the past decade at facilities such as JLab/CLAS, MAMI, and MIT-Bates.
        Experimental data exist now up to $Q^2 \sim 8$ GeV$^2$;
        for comprehensive reviews on the present experimental status we refer to Refs.~\cite{Aznauryan:2011qj,Pascalutsa:2006up}.

        While it contributes less than $1\%$ to the total decay width of the Delta-isobar,
        the $N \to \Delta \gamma$ transition represents an indubitable test of the
        deformation from sphericity in both nucleon and $\Delta$ baryons.
        It is characterized by three transition form factors: the magnetic dipole form factor $M1$,
        the electric quadrupole form factor $E2$, and the Coulomb quadrupole form factor $C2$,
        which can be equivalently expressed in terms of helicity amplitudes.
        Although the magnetic dipole transition is dominant, an accurate extraction of the form factor
        ratios $R_{EM}=E2/M1$ and $R_{SM}=C2/M1$ reveals small and negative values.
        This constitutes a measure of the non-spherical distribution of the partons
        within the baryons involved in the reaction.

        The theoretical description of the $N\to\Delta\gamma$ transition has been traditionally quite challenging.
        In a constituent-quark model where three quarks move non-relativistically in an $s$ wave, the transition magnetic moment is underestimated by $25\%$, and
        the electromagnetic ratios are only non-zero if $d$ waves are included~\cite{Becchi:1965zz,Isgur:1981yz}.
        More sophisticated approaches have been developed over the years with varying degrees of improvement.
        The $N\to\Delta\gamma$ transition has been studied in chiral effective field theory~\cite{Pascalutsa:2005vq,Gail:2005gz},
        dynamical reaction models such as the Sato-Lee model~\cite{Sato:2000jf,JuliaDiaz:2006xt},
        light-cone QCD sum-rule analyses~\cite{Lenz:2009ar},
        large-$N_C$ approaches~\cite{Pascalutsa:2007wz, Grigoryan:2009pp},
        the general parametrization method~\cite{Dillon:2009pf},
        the cloudy bag model~\cite{Bermuth:1988ms,Lu:1996rj},
        Skyrme models~\cite{Abada:1995db, Wirzba:1986sc, Walliser:1996ps},
        vector-meson-dominance types of models~\cite{Vereshkov:2007xy},
        relativistic quark models improved by chiral corrections~\cite{Faessler:2006ky},
        and other constituent quark models~\cite{Hemmert:1994ky, Buchmann:2004ia, Ramalho:2008aa, Ramalho:2008dp}.
        Systematic progress in recent years has also been made by calculations in lattice QCD~\cite{Alexandrou:2007dt,Alexandrou:2010uk}.

        The nucleon and $\Delta$ deformation is tightly connected to the understanding of their internal structure.
        In Poincar\'e-covariant approaches, the source of deformation from sphericity is attributed to the
        quark orbital angular momentum distribution among a hadron's constituents, predominantly arising from $p$ waves
        in the hadron amplitudes, and to a much lesser extent from $d$ waves~\cite{Oettel:1998bk,Nicmorus:2010sd,Eichmann:2011vu}.
        In addition, the properties of the nucleon and its first resonance are sensitive to chiral corrections and expected
        to be generated by an interplay between pure quark states and pionic clouds.

        In support of this, a QCD-motivated analysis of the quark-core contribution to the $N \to \Delta \gamma$ transition properties is desirable.
        We employ the framework of Dyson-Schwinger equations (DSEs) and Bethe-Salpeter equations (BSEs) which provides a well-established tool of investigation.
        In the quantum field theory of color-charged particles, dynamical chiral symmetry breaking, confinement and
        the formation of hadron bound states are phenomena which require a non-perturbative treatment.
        DSEs represent a fully self-consistent infinite set of coupled integral equations for QCD's Green functions.
        The Dyson-Schwinger framework is particularly suitable to address these phenomena as it provides systematic access to both perturbative and non-perturbative regimes of QCD;
        see~\cite{Roberts:1994dr,Alkofer:2000wg,Fischer:2006ub} for reviews.

        In the present approach, hadrons are studied via covariant bound-state equations,
        see~\cite{Maris:2003vk, Roberts:2007jh, Eichmann:2009zx} and references therein. Meson properties emerge from solutions of the $q\bar{q}$ Bethe-Salpeter equation,
        while baryon properties are obtained from its three-body equivalent, the covariant Faddeev equation.
        Although of far bigger complexity, the Faddeev equation was recently solved for the nucleon and $\Delta-$baryon masses~\cite{Eichmann:2009qa,Eichmann:2009en,SanchisAlepuz:2011jn}
        and utilized to calculate the nucleon's electromagnetic, axial and pseudoscalar form factors~\cite{Eichmann:2011vu,Eichmann:2011pv}.
        By implementing a rainbow-ladder (RL) truncation,
        i.e. a dressed gluon-exchange kernel between any two quarks,
        these investigations enable a direct comparison with corresponding meson studies.
        Meson-cloud effects in the chiral and low-momentum structure of hadrons
        are presently not accounted for, hence the framework aims at a description of the hadronic quark core.

        It is interesting that the nucleon and $\Delta$ masses and nucleon form factors
        obtained from three-body Faddeev calculations show little discrepancies to those obtained in
        a simplified version of the approach, the quark-diquark model.
        The latter represents an efficacious simplification of the Faddeev three-body problem to a
        Bethe-Salpeter two-body problem, where correlations in the $qq$ scattering matrix
        beyond the dominant scalar and axial-vector diquarks are neglected~\cite{Oettel:1998bk,Eichmann:2007nn}.
        The quark-diquark approach is based on the observation that the attractive nature of quark-antiquark
        correlations in a color-singlet meson is also attractive for $\bar{3}_C$ quark-quark correlations
        within a color-singlet baryon.
        In this respect, we consider a first study of the $N \to \Delta \gamma $ transition properties within a consistent and well-established quark-diquark framework worthwhile
        and therefore adopt this simplification in the present paper.
        Moreover, this study completes previous investigations of quark-core contributions to the nucleon and $\Delta-$baryon masses
        and form factors~\cite{Eichmann:2008ef,Eichmann:2008kk,Nicmorus:2008vb,Nicmorus:2008eh,Mader:2011zf},
        where the backbone of the approach was entirely provided by the quark-diquark model.

        The manuscript is organized as follows: in Section~\ref{sec:Faddeev} we discuss the Poincar\'e-covariant
        Faddeev approach to baryons and its quark-diquark simplified setup.
        In Section~\ref{sec:ffs} we elaborate on the construction and properties of the electromagnetic transition current operator
        in the quark-diquark framework. In Section~\ref{sec:results} we present and comment on the results for
        the $N \to \Delta \gamma$ transition form factors and compare them to experimental data.
        Technical details of our calculations are presented in Appendices~\ref{app:conventions}--\ref{app:current-details}.
        We work in Euclidean momentum space and use the isospin-symmetric limit $m_u=m_d$.

%%%%%%%%%%%%%%%%%%%%%%%%%%%%%%%%%%%%%%%%%%%%%%%%%%%%%%%%%%%%%%%%%%%%%%%%%%%%%%%%%%%%%%%%%%%%%%%%%%%%%%%%%%%%%%%%%%%%%%%%%%%%

\section{Quark-diquark framework} \label{sec:Faddeev}

          The description of the $N\Delta\gamma$ transition properties in the Dyson-Schwinger approach requires knowledge of
          the nucleon and $\Delta$ bound-state amplitudes and their microscopic ingredients in terms of QCD's Green functions.
          The quantities which appear explicitly in the computation of hadron wave functions and form factors
          are the dressed quark propagator and the irreducible $q\bar{q}$, $qq$ and $qqq$ kernels.
          They encode the interactions at the quark-gluon level and, at least in principle, can be obtained through
          QCD's Dyson-Schwinger equations.

          In order to compute nucleon and $\Delta$ properties,
          these building blocks must be combined via covariant bound-state equations.
          In the present study we treat baryons as bound states of quarks and diquarks which amounts to a simplification of the three-quark problem.
          Irreducible three-quark interactions are neglected, and the $qq$ interaction is subsumed in effective diquark correlations.
          As a consequence, gluons appear only implicitly in quark and diquark propagators and quark-diquark vertex functions.

          In subsection~\ref{sec:quarks-diquarks} we will briefly
          describe the basic input of the approach, namely, the rainbow-ladder ansatz
          for the $qq$ kernel which amounts to an iterated gluon exchange between two quarks.
          The subsequent calculation of diquark and baryon properties from that input
          is discussed in~\ref{sec:diquarks} and~\ref{sec:amplitudes}, respectively.

\subsection{Quarks} \label{sec:quarks-diquarks}

           The basic building blocks that enter the description of hadronic bound states
           are the dressed quark propagator $S(p)$ and the two-quark irreducible kernel $\mathcal{K}$.
           These are the ingredients that appear in a meson's Bethe-Salpeter equation and the covariant Faddeev equation for a baryon.
           The quark propagator satisfies a Dyson-Schwinger equation which can be solved numerically within a suitable truncation.
           On the other hand, the absence of a self-consistent solution for the quark four-point function necessitates an ansatz for the kernel $\mathcal{K}$.

           The construction of kernel ans\"atze is guided by vector and axialvector Ward-Takahashi identities
           which ensure electromagnetic current conservation as well as Gell-Mann-Oakes-Renner and Goldberger-Treiman relations at the hadron level~\cite{Maris:1997hd,Holl:2004fr,Eichmann:2011pv}.
           These identities impose constraints on the structure of the kernel $\mathcal{K}$ by relating it with the kernel of the quark DSE~\cite{Munczek:1994zz,Bender:1996bb}.
           The simplest ansatz to satisfy those constraints is the rainbow-ladder kernel:
                \begin{equation}\label{RLkernel}
                    \mathcal{K}_{\alpha\alpha'\beta\beta'} =  Z_2^2 \, \frac{ 4\pi \alpha(k^2)}{k^2} \,
                    T^{\mu\nu}_k \gamma^\mu_{\alpha\alpha'} \,\gamma^\nu_{\beta\beta'},
                \end{equation}
           where $T^{\mu\nu}_k=\delta^{\mu\nu} - k^\mu k^\nu/k^2$
           is a transverse projector with respect to the gluon momentum $k$, and $Z_{2}$ is the quark renormalization constant.
           Eq.~\eqref{RLkernel} describes a dressed gluon exchange between quark and antiquark that retains
           only the vector part $\sim \gamma^\mu$ of the quark-gluon vertex.
           Its non-perturbative dressing, together with that of the gluon propagator,
           is absorbed by an effective coupling $\alpha(k^2)$ which is modeled.

             \begin{figure}[t]
                        \begin{center}

                        \includegraphics[scale=0.13]{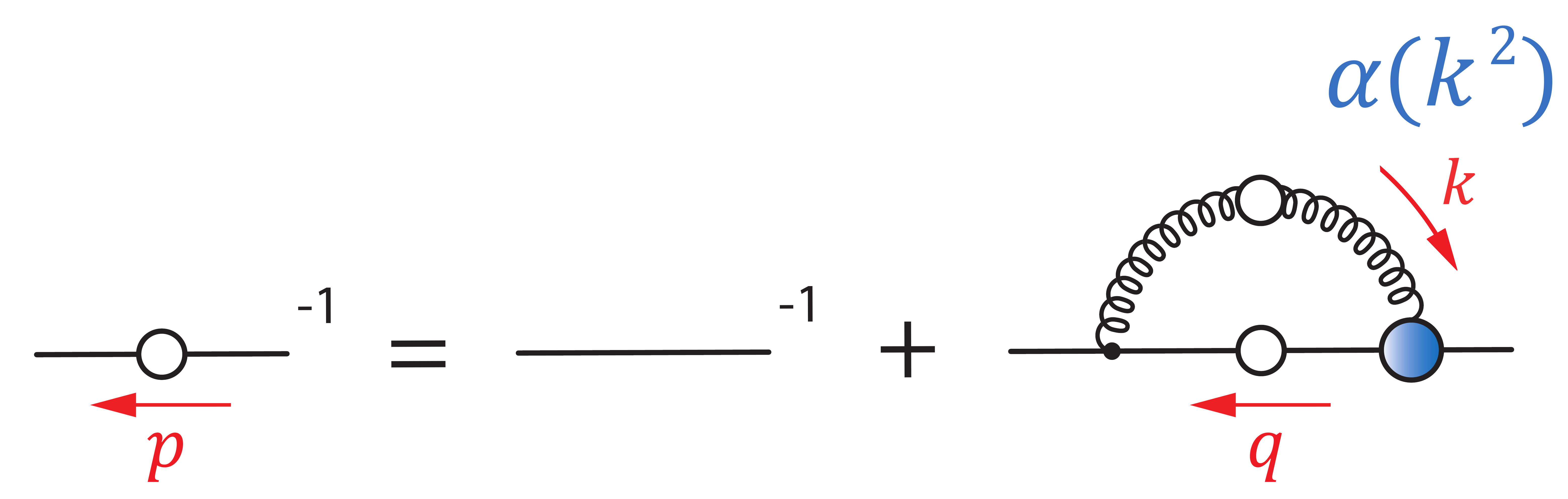}
                        \caption{(Color online) Quark DSE~\eqref{quarkdse} in rainbow-ladder truncation.
                                 }\label{fig:dse}

                        \end{center}
             \end{figure}

           The second basic ingredient is the dressed quark propagator.
           It is expressed by two scalar functions, the quark wave-function renormalization $1/A(p^2)$ and the quark mass function $M(p^2)$:
                  \begin{equation}\label{qprop}
                      S^{-1}(p) = A(p^2)\,\left( i\Slash{p} + M(p^2) \right).
                  \end{equation}
           The quark propagator satisfies the quark DSE which is illustrated in Fig.~\ref{fig:dse}.
           Its interaction kernel includes the dressed gluon propagator as well as one bare
      	   and one dressed quark-gluon vertex.
           In rainbow-ladder truncation that kernel becomes identical to Eq.~\eqref{RLkernel} and the quark DSE reads:
                  \begin{equation}\label{quarkdse}
                      S^{-1}_{\alpha\beta}(p) = Z_2 \left( i\Slash{p} + m_0 \right)_{\alpha\beta}  +
      		               \int \mathcal{K}_{\alpha\alpha'\beta'\beta} \,S_{\alpha'\beta'}(q)\,.
                  \end{equation}
           The bare current-quark mass $m_0$
           appears here as the input of the equation and can be varied from the chiral limit up to the heavy-quark regime.

\begin{figure}[tbp]
                    \begin{center}

                    \includegraphics[scale=0.40]{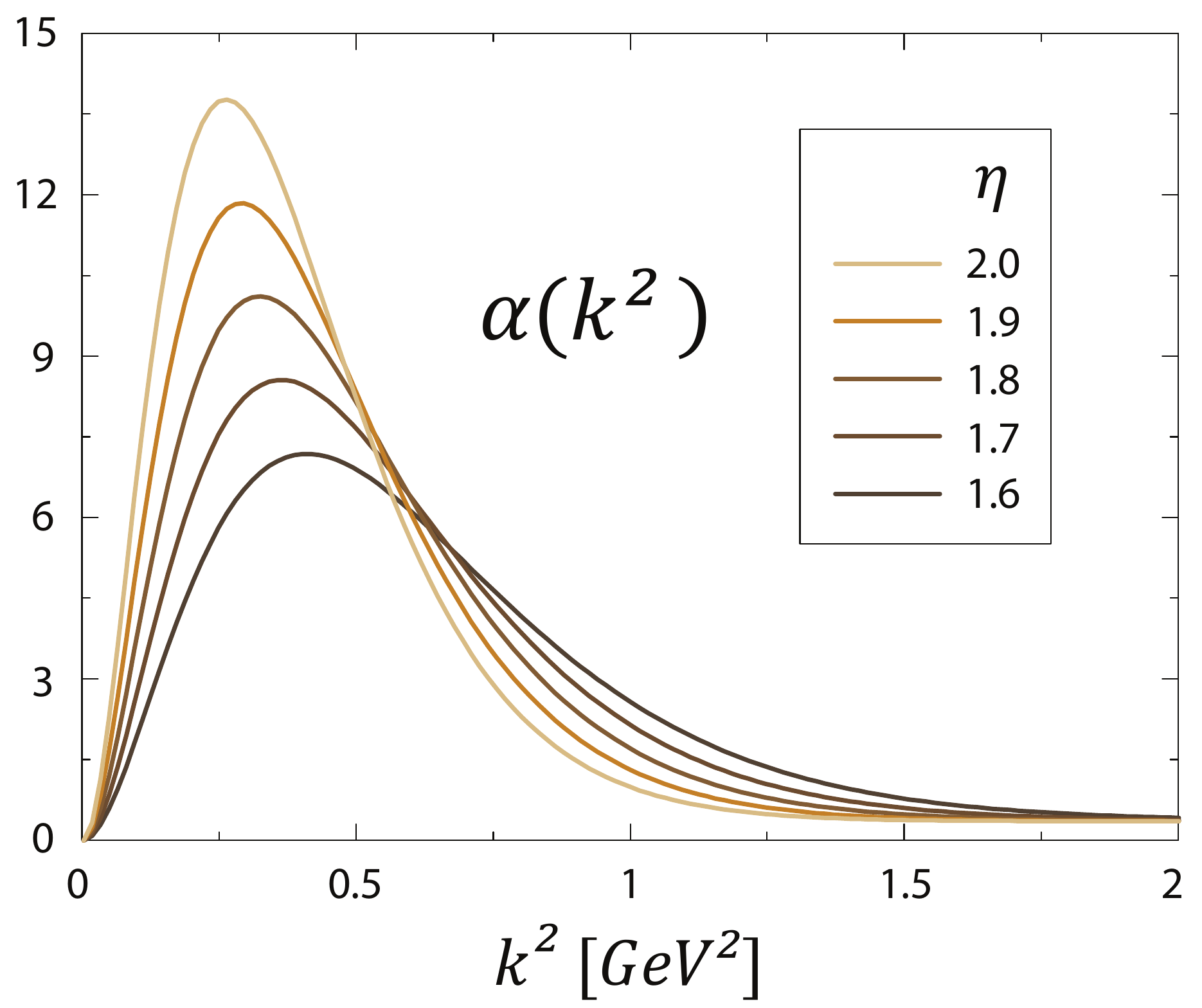}
                    \caption{(Color online) Effective coupling $\alpha(k^2)$ of Eq.\,\eqref{couplingMT}, evaluated for $\Lambda = 0.72$ GeV
                                            and in the range $\eta \in [1.6,2.0]$.}\label{fig:coupling}

                    \end{center}
        \end{figure}

           The quark DSE exhibits dynamical chiral symmetry breaking if the kernel $\mathcal{K}$ supplies sufficient interaction strength.
           Its consequence is a non-perturbative enhancement of the quark mass function $M(p^2)$ in the low-momentum region that
           indicates the dynamical generation of a constituent-quark mass scale.
           In principle, such strength would be generated through a self-consistent DSE solution for the gluon propagator and quark-gluon vertex
           that enter the quark DSE.
           In the rainbow-ladder truncation, that effect is provided by the effective coupling $\alpha(k^2)$ which appears in the kernel~\eqref{RLkernel}.
           The ansatz we choose is taken from Ref.~\cite{Maris:1999nt} and plotted in Fig.~\ref{fig:coupling}:
                        \begin{equation}\label{couplingMT}
                            \alpha(k^2) = \pi \eta^7  \left[\frac{k^2}{\Lambda^2}\right]^2 \!\!
                            e^{-\eta^2 \left[\frac{k^2}{\Lambda^2}\right]} + \alpha_\text{UV}(k^2) \,.
                        \end{equation}
           The second term $\alpha_\text{UV}$ is only relevant at large gluon momenta where it dominates and is constrained by perturbative QCD.
           The important part in view of hadron properties is the first term. It provides the necessary strength at small and intermediate momenta
           that triggers the transition from a current-quark to a dynamically generated constituent quark.
           It is characterized by two parameters:
           an infrared scale $\Lambda$ that represents the scale of dynamical chiral symmetry breaking,
           and a dimensionless width parameter $\eta$ that modifies the shape of the interaction in the infrared, cf. Fig.~\ref{fig:coupling}.

           In combination with the interaction of Eq.~\eqref{couplingMT}, the rainbow-ladder truncation
           has been quite successful in describing a variety of hadron properties.
           Upon setting the scale $\Lambda$ via the experimental pion decay constant,
           its implementation in hadronic bound-state equations describes pseudoscalar-meson, vector-meson,
           nucleon and $\Delta$ ground-state properties reasonably well, see~\cite{Maris:2005tt,Maris:2006ea,Krassnigg:2009zh,Nicmorus:2010mc,Eichmann:2011vu,SanchisAlepuz:2011jn} and references therein.
           In addition, their features show an overall insensitivity to a variation of the infrared properties of the
           coupling which is controlled by the parameter $\eta$~\cite{Maris:1999nt,Krassnigg:2009zh}.

           Progress in the light meson sector has also been made for axial-vector and pseudoscalar isosinglet mesons.
           Their properties are subject to substantial corrections beyond rainbow-ladder which mainly
           come from the quark-gluon vertex~\cite{Alkofer:2008et,Fischer:2009jm,Chang:2011vu}.
           Another important type of corrections are
           pion-cloud contributions which are important in the chiral and low-momentum structure of hadrons.
           They give rise to chiral singularities and non-analyticities associated with the opening of decay channels.
           Pion-cloud effects are not implemented in the present approach which therefore
           aims at investigating the properties of the nucleon and $\Delta$ quark core.

           Eq.~\eqref{couplingMT} remains the basic model input throughout this work.
           By setting the scale $\Lambda$ and treating $\eta$ as a parameter that reflects the model uncertainty,
           the properties of the effective interaction directly translate to the level of form factors.
           The quantities that appear in intermediate steps are computed self-consistently
           from this input so that no further model assumptions are necessary.

        \subsection{Diquarks} \label{sec:diquarks}

         The importance of diquark correlations in view of understanding baryon structure
         has a longstanding history~\cite{Anselmino:1992vg,Jaffe:2004ph}.
         The strong attraction in the color anti-triplet diquark channel has motivated the interpretation of
         baryons as bound states of quarks and scalar ('good') diquarks. Such a picture reduces the rotational degrees of freedom in baryons and
         can explain the absence of baryon resonances that are predicted by symmetric quark models.
         Nevertheless, recent lattice results for excited baryon masses show more similarity with traditional quark model predictions~\cite{Edwards:2011jj}.

         In that respect, one should be cautious in order not to oversimplify the problem.
         Diquark correlations certainly carry internal structure, and the inclusion of axialvector diquarks is crucial as well~\cite{Ishii:1995bu,Oettel:1998bk}.
         In our setup, diquarks approximate the structure of the $qq$ scattering matrix $T$ that enters a baryon's Faddeev equation.
         It is expressed as the separable sum of scalar and axialvector-diquark contributions:
         \begin{equation}\label{Tmatrix-dq}
         \begin{split}
             T_{\alpha\beta\beta'\alpha'} &\approx \sum_\text{D} \Gamma^\mu_\text{D}(p,P)_{\alpha\beta}\,D^{\mu\nu}(P^2)\,\conjg{\Gamma}^\nu_\text{D}(p',P)_{\beta'\alpha'}\,,
         \end{split}
         \end{equation}
         whose ingredients, in terms of diquark propagators $D^{\mu\nu}$ and quark-diquark vertices $\Gamma_\text{D}^\mu$, are computed dynamically from QCD.
         Here, the label 'D' denotes the type of diquark. For scalar diquarks one has $\mu=\nu=0$ and for axialvector diquarks $\mu,\nu=1\dots 4$.
         The total $qq$ momentum is called $P$, and $p$, $p'$ are the relative momenta between the quarks.
         A bar on an amplitude denotes charge conjugation, cf.~App.~\ref{app:conventions}.

         The implementation of the full $T-$matrix in a baryon's bound-state equation is equivalent
         to solving its covariant Faddeev equation without irreducible three-quark interactions, as implemented in Refs.~\cite{Eichmann:2009qa,SanchisAlepuz:2011jn}.
         From that point of view, the quark-diquark model defined by Eq.~\eqref{Tmatrix-dq} merely amounts to a truncation of the Faddeev equation.
         In fact, both approaches yield quite similar results for nucleon and $\Delta$ masses
         and nucleon electromagnetic form factors~\cite{Nicmorus:2010mc,Eichmann:2011vu,SanchisAlepuz:2011jn},
         which indicates that scalar and axialvector diquarks provide indeed the overwhelming contribution to the binding of these baryons.
         The question of baryon excitations, on the other hand, has not yet been addressed in full sophistication in either of these frameworks.
         A recent study of the Roper resonance in a simpler model suggests a quark-diquark structure as well~\cite{Roberts:2011ym}.

         While the dominance of $qq$ correlations in the structure of baryons may be a generic feature,
         the presence of explicit \textit{timelike} diquark poles in the T-matrix
         is tied to the properties of the rainbow-ladder kernel~\eqref{RLkernel}.
         The rainbow-ladder truncation generates $qq$ poles in the full $T-$matrix which are recovered by the ansatz~\eqref{Tmatrix-dq}
         in terms of scalar and axialvector diquark propagators:
         \begin{equation}\label{dq-pole}
         \begin{split}
             D^{00}(P^2) &\stackrel{P^2\rightarrow -m_\text{sc}^2}{\longlonglongrightarrow} \frac{1}{P^2+m_\text{sc}^2}\,, \\
             D^{\mu\nu}(P^2) &\stackrel{P^2\rightarrow -m_\text{av}^2}{\longlonglongrightarrow} \frac{T_P^{\mu\nu}}{P^2+m_\text{av}^2}\,,
         \end{split}
         \end{equation}
         where $T_P^{\mu\nu}$ is again the transverse projector defined below Eq.~\eqref{RLkernel}.
         Colored diquarks disappear from the physical spectrum beyond rainbow-ladder; hence, these poles are truncation artifacts~\cite{Bender:1996bb}.
         Nevertheless, they provide access to a dynamical computation of onshell diquark amplitudes and their masses $m_\text{sc}$, $m_\text{av}$ from the diquark BSE:
         \begin{equation}\label{diquark-bse}
         \begin{split}
             &\Gamma^\mu_\text{D}(p,P)_{\alpha\beta}= \\
             &=\int\limits_q \mathcal{K}_{\alpha\alpha'\beta'\beta}\, \left[S(q_+)\,\Gamma^\mu_\text{D}(q,P)\,S^T(-q_-)\right]_{\alpha'\beta'}\,,
         \end{split}
         \end{equation}
         with $q_\pm = q \pm P/2$. Eq.~\eqref{diquark-bse} has a similar structure as the BSEs for pseudoscalar and vector mesons: upon tracing out the Dirac structure,
         they only differ by a color factor. Details about the Lorentz-Dirac structure of the diquark amplitudes and the solution method
         can be found in Refs.~\cite{Maris:2002yu,Eichmann:2009zx}.

         Diquark correlations in a baryon are offshell. The offshell structure of the $qq$ scattering matrix in the separable approximation is primarily
         encoded in the diquark propagators. Inserting Eq.~\eqref{Tmatrix-dq} in the scattering equation for the $T-$matrix yields an expression
         that allows for a consistent determination of the propagators. It reads schematically~\cite{Eichmann:2009zx}:
         \begin{equation}\label{diquark-propagator}
         D^{-1}(P^2)  = \text{Tr}\int\conjg{\Gamma}_D  \,(G_0-\mathcal{K}^{-1})\,\Gamma_D    \,,
         \end{equation}
         where $G_0$ is the disconnected product of two quark propagators and $\mathcal{K}$ the rainbow-ladder kernel.
         It involves the onshell diquark amplitudes from Eq.~\eqref{diquark-bse} that must be equipped with appropriate offshell ans\"atze.
         The resulting diquark propagators reproduce Eq.~\eqref{dq-pole} on the respective mass shells
         but deviate from free-particle propagators at offshell momenta.

         \begin{figure}[t]
                    \begin{center}

                    \includegraphics[scale=0.34]{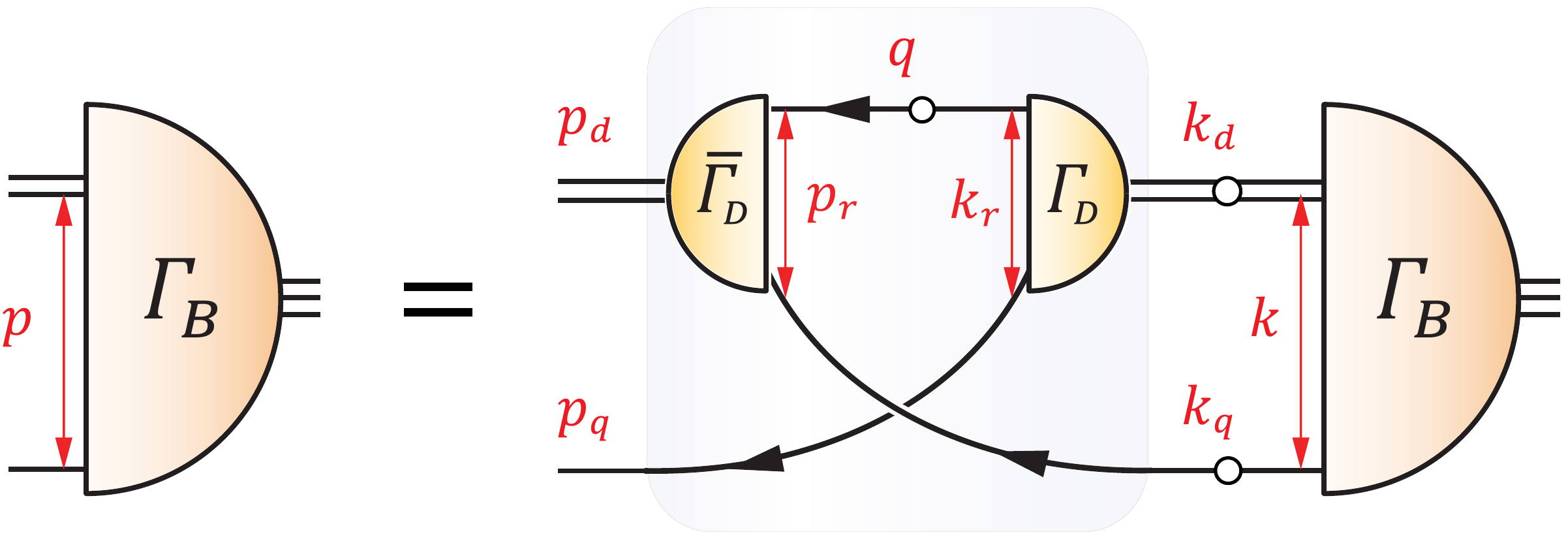}
                    \caption{(Color online) The quark-diquark BSE, Eq.~\eqref{quark-diquark-bse}.\label{fig:qdq-BSE}}

                    \end{center}
        \end{figure}

\subsection{Nucleon and $\Delta$ bound-state equations} \label{sec:bses}

         The simplification of the three-body problem to an effective quark-diquark interpretation of baryons
         proceeds via omitting genuine three-quark interactions and employing the separable pole expansion~\eqref{Tmatrix-dq}
         in the quark-quark $T$-matrix.
         The resulting interaction between quark and diquark is an iterative quark exchange,
         where in every iteration step the spectator quark and one quark inside the
         diquark exchange roles~\cite{Oettel:1998bk,Oettel:2000jj,Eichmann:2007nn,Eichmann:2009zx}.

         The corresponding quark-diquark BSE is illustrated in Fig.~\ref{fig:qdq-BSE} and reads
         \begin{equation}\label{quark-diquark-bse}
          \Gamma_\text{B}^\alpha(p,P)= \int\limits_k K_\text{Q-DQ}^{\alpha\beta}\, S(k_q) D^{\beta\beta'}(k_d)\,\Gamma_\text{B}^{\beta'}(k,P)\, ,
         \end{equation}
         where $\Gamma_\text{B}^\alpha$ are the quark-diquark amplitudes of the respective baryon, with B = $N$ or $\Delta$.
         The quark-diquark exchange kernel is given by
         \begin{equation}\label{quark-diquark-kernel}
             K_\text{Q-DQ}^{\alpha\beta} = \Gamma_\text{D}^\beta(k_r,k_d)\,S^T(q)\,\conjg{\Gamma}_\text{D}^\alpha(p_r,p_d)\,.
         \end{equation}
         The scalar and axialvector diquark amplitudes and propagators that appear in these expressions were explained in connection with Eq.~\eqref{Tmatrix-dq}.
         $p_{q,d}$ and $k_{q,d}$ are the external and internal quark and diquark momenta, $p_r$ and $k_r$ are
         the relative momenta that enter the diquark amplitudes, and $P$ is here the total baryon momentum, cf.~Fig.~\ref{fig:qdq-BSE}.
         The Dirac-Lorentz structure of the baryon amplitudes $\Gamma_\text{B}^\alpha$ is analyzed in the following subsection.
         For details on the solution of the quark-diquark BSE we refer to Refs.~\cite{Oettel:2001kd,Eichmann:2009zx}.

         The treatment of light baryons such as the nucleon and $\Delta$ in the quark-diquark approach usually
         retains the lightest diquark degrees of freedom, i.e., scalar and axial-vector diquarks.
         The $\Delta$ baryon then involves only axial-vector diquark correlations whereas the nucleon contains both.
         As a consequence, the $N\Delta\gamma$ transition will not only involve axial-scalar diquark transitions
         but also axial-axial correlations. We will return to the implications of this feature in Section~\ref{sec:results-Q2}.

\subsection{Nucleon and $\Delta$ amplitudes} \label{sec:amplitudes}

        When expressed as two-body bound states of quarks and diquarks,
        the Dirac-Lorentz structure of the nucleon and $\Delta$ amplitudes $\Gamma_\text{B}^\alpha$
        is considerably simpler compared to the full three-body approach of Refs.~\cite{Eichmann:2009qa,SanchisAlepuz:2011jn}.
        Their derivation and partial-wave decomposition in the quark-diquark model
        has been outlined in Refs.~\cite{Oettel:1998bk,Oettel:2000ig}. In view of exploring the impact of different
        quark orbital-angular momentum eigenstates on the properties of the $N\Delta\gamma$ transition,
        we briefly repeat the construction here for convenience.

            \setlength{\extrarowheight}{2pt}

            \begin{table}
                \begin{center}

                \renewcommand{\arraystretch}{1.4}

                \begin{equation*}
                \begin{array}{ | @{\quad}  c @{\quad} |  @{\;\;}l@{\;\;} |     }\hline

                   \Gamma_N^0  &
                   \begin{array}{ @{\quad}l @{\quad}}
                       \tau_1 = \mathds{1} \\
                       \tau_2 = i\Slash{r} \\[0.2cm]
                   \end{array}    \\ \hline

                   \Gamma_N^\mu &
                   \begin{array}{ @{\quad} l @{\quad}}
                       \tau_3^\mu = \frac{1}{\sqrt{3}}\,\gamma^\mu_T                \\
                       \tau_4^\mu = \frac{1}{\sqrt{3}}\,\gamma^\mu_T\,i\Slash{r}    \\
                       \tau_5^\mu = \hat{P}^\mu                                      \\
                       \tau_6^\mu = \hat{P}^\mu \,i\Slash{r}                         \\
                       \tau_7^\mu = \frac{1}{\sqrt{6}}\left( \gamma^\mu_T - 3 r^\mu \Slash{r}\right)   \\
                       \tau_8^\mu = \frac{1}{\sqrt{6}}\left( \gamma^\mu_T - 3 r^\mu \Slash{r}\right) i\Slash{r} \\[0.2cm]
                   \end{array}   \\ \hline

                   \Gamma_\Delta^{\mu\nu} &
                   \begin{array}{ @{\quad} l @{\quad}}
                       \tau_1^{\mu\nu} = \delta^{\mu\nu}  \\
                       \tau_2^{\mu\nu} = \textstyle\frac{i}{\sqrt{5}}\left( 2 \,\gamma_T^\mu \,r^\nu - 3 \,\delta^{\mu\nu} \Slash{r}\right)  \\
                       \tau_3^{\mu\nu} =  \sqrt{3}\,\hat{P}^\mu \,r^\nu \Slash{r}   \\
                       \tau_4^{\mu\nu} = \sqrt{3}\,\hat{P}^\mu \,ir^\nu            \\
                       \tau_5^{\mu\nu} = \gamma_T^\mu \,r^\nu \Slash{r}     \\
                       \tau_6^{\mu\nu} = \gamma_T^\mu \,ir^\nu   \\
                       \tau_7^{\mu\nu}  = -\gamma_T^\mu \,r^\nu \Slash{r} - \delta^{\mu\nu} + 3 \,r^\mu r^\nu \\
                       \tau_8^{\mu\nu}  = \textstyle\frac{i}{\sqrt{5}}\left( \delta^{\mu\nu} \Slash{r} + \gamma_T^\mu \,r^\nu - 5 \,r^\mu r^\nu \Slash{r}\right)  \\[0.2cm]
                   \end{array}   \\ \hline

                \end{array}
                \end{equation*}

                \caption{Orthonormal basis elements for the nucleon and $\Delta$ quark-diquark amplitudes that appear in Eqs.~(\ref{nuc:amplitudes}--\ref{delta:amplitudes}).  }\label{table:basis}

                \renewcommand{\arraystretch}{1.0}
                \end{center}
            \end{table}

             \begin{table}[t]
                        \begin{center}
                        \includegraphics[scale=0.73]{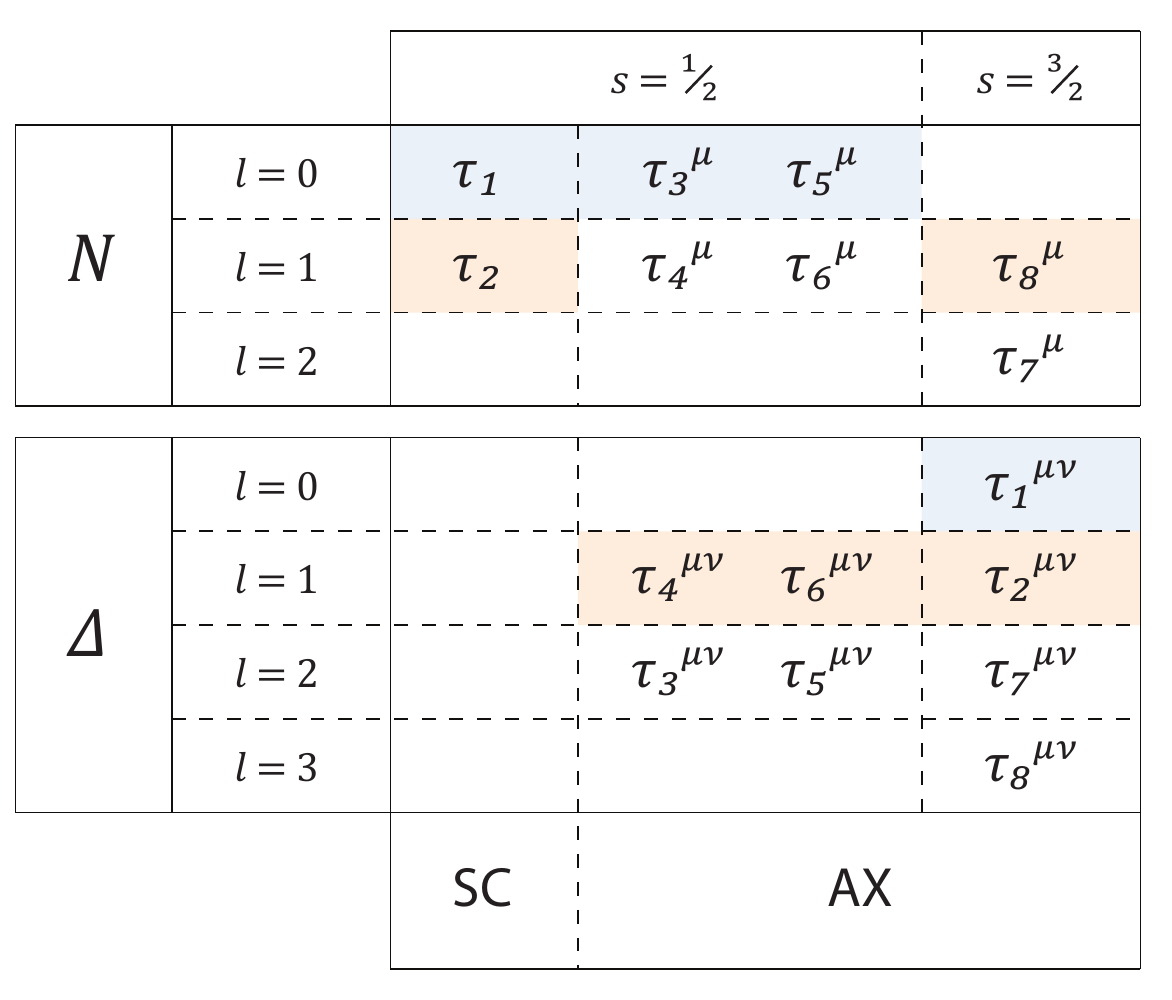}
                        \caption{(Color online) Rest-frame partial-wave decomposition of nucleon and $\Delta-$baryon
                                                in the quark-diquark approach. The basis elements are characterized
                                                by their scalar and axialvector diquark content and
                                                their eigenvalues with respect to quark-diquark spin ($s$) and orbital
                                                angular momentum ($l$). The colored boxes highlight the dominant components;
                                                for example, the dressing functions associated with $\tau_4^\mu$ and $\tau_6^\mu$
                                                are much smaller than the remaining $p-$wave contributions.
                                                 }\label{fig:partial-wave}
                        \end{center}
            \end{table}

            A two-body system is described by two independent momenta, the total momentum $P$ and the relative momentum $p$.
            Three Lorentz-invariant momentum variables can be constructed out of those: $p^2$, $z:=\hat{p}\cdot\hat{P}$, and $P^2$.
            The total momentum is onshell: $P^2=-M_\text{B}^2$, where the index B denotes the nucleon or $\Delta-$baryon.
            The quark-diquark amplitudes can be expanded in a basis whose dimension is determined
            by Poincar\'e covariance and parity invariance,
            together with the restrictions that follow if the baryon is taken on its mass shell.
            The respective decomposition for the scalar and axial-vector diquark parts of the nucleon amplitude reads:
            \begin{equation}\label{nuc:amplitudes}
                \begin{split}
                    \Gamma_N^0(p,P) &=   \sum_{k=1}^2 f_k^N(p^2,z) \,\tau_k(r,\hat{P}) \, \Lambda_+(P)\,, \\
                    \Gamma_N^\mu(p,P) &= \sum_{k=3}^8 f_k^N(p^2,z) \,\tau_k^\mu(r,\hat{P}) \, \gamma^5 \Lambda_+(P)\,,  \\
                \end{split}
            \end{equation}
            and for the $\Delta$ amplitude:
            \begin{equation}\label{delta:amplitudes}
                    \Gamma_\Delta^{\mu\nu}(p,P) = \sum_{k=1}^8 f_k^\Delta(p^2,z) \, \tau_k^{\mu \rho}(r,\hat{P})\, \mathds{P}^{\rho \nu}(P)\,.
            \end{equation}
            The Lorentz-invariant coefficients $f_k^\text{B}$ depend on the two invariant variables $p^2$ and $z$
            and are solutions of the quark-diquark bound-state equations.
            We have factored out the nucleon and $\Delta$ spinors  and work
            instead with the positive-energy and Rarita-Schwinger projectors which satisfy
             \begin{equation}\label{spinors}
             \begin{split}
                 \Lambda_+(P)\,u(P,s) &= u(P,s)\,, \\
                 \mathds{P}^{\mu\nu}(P) u^\nu(P,s) &= u^\mu(P,s)\,.
             \end{split}
             \end{equation}
             They are given by
             \begin{equation}\label{projectors}
                 \Lambda_+ = \textstyle\frac{1}{2}\,(\mathds{1}+\hat{\Slash{P}})\,,\quad
                 \mathds{P}^{\mu\nu}= \Lambda_+ \Big(T_{P}^{\mu\nu}-\textstyle\frac{1}{3}\,\gamma^\mu_T \,\gamma^\nu_T\Big),
             \end{equation}
             where $\hat{P}=P/(iM_\text{B})$ is the normalized baryon momentum,
             $T_P^{\mu\nu} = \delta^{\mu\nu} - \hat{P}^\mu\hat{P}^\nu$ is a transverse projector with respect to $P$,
             and $\gamma^\mu_T = T_P^{\mu\nu} \gamma^\nu$ is the transverse $\gamma-$matrix.
             The projectors inherit the constraints from the spinors:
             \begin{equation}\label{projector-contractions}
                 \hat{\Slash{P}}\,\Lambda_+= \Lambda_+\,, \quad
                 \hat{P}^\mu \,\mathds{P}^{\mu\nu}= \gamma^\mu \,\mathds{P}^{\mu\nu} = 0\,.
             \end{equation}

            Instead of $p$ and $P$, the basis elements in Eqs.~(\ref{nuc:amplitudes}--\ref{delta:amplitudes}) can be equally well expressed through
            orthonormal momenta $\hat{P}^\mu$ and $r^\mu := \widehat{p_T}^\mu $, i.e., such that $r^2=\hat{P}^2=1$ and $r\cdot \hat{P}=0$.
            The dependence on the Lorentz invariants $p^2$ and $z$ is then carried by the coefficients $f_k^\text{B}$ only.
            This simplifies the construction of an orthogonal basis and is also convenient for practical calculations,
            e.g. in the baryon's rest frame, where $\hat{P}$ and $r$ are Euclidean unit vectors.

            The largest linearly independent set of basis elements for the bound-state amplitude $\Gamma_N^0$, $\Gamma_N^\mu$ and
            $\Gamma_\Delta^{\mu\nu}$ is given in Eq.~\eqref{qdq-basis-general}. On the baryon's mass shell,
            which is enforced by the properties~\eqref{projector-contractions} of the projectors, the following independent basis elements remain:
            \begin{equation} \label{qdq-basis-onshell}
            \begin{split}
            \Gamma_N^0: & \quad \{ \mathds{1}, \, \Slash{r} \}, \\
            \Gamma_N^\mu: & \quad  \{ \gamma_T^\mu, \,r^\mu, \,\hat{P}^\mu \} \times \{ \mathds{1}, \, \Slash{r} \},\\
            \Gamma_\Delta^{\mu\nu}:  & \quad   \{ \delta^{\mu\nu}, \,\gamma_T^\mu \,r^\nu, \,r^\mu r^\nu, \,\hat{P}^\mu r^\nu \} \times \{ \mathds{1}, \, \Slash{r} \}.
            \end{split}
            \end{equation}
            These can be further orthonormalized and arranged according to
            their (quark-diquark) spin and orbital angular momentum content in the baryon's rest frame, cf.~App.~\ref{app:partial-wave}.
            The resulting classification in $s$, $p$, $d$ and $f$ waves is illustrated in Tables~\ref{table:basis} and~\ref{fig:partial-wave}.
            We emphasize that $p$-wave contributions to the bound-state amplitudes emerge quite naturally because of Poincar\'e covariance.
            Those disappear in the non-relativistic limit~\cite{Oettel:2000ig}
            but have important consequences for the behavior of the form factors in Section~\ref{sec:results}.

%%%%%%%%%%%%%%%%%%%%%%%%%%%%%%%%%%%%%%%%%%%%%%%%%%%%%%%%%%%%%%%%%%%%%%%%%%%%%%%%%%%%%%%%%%%%%%%%%%%%%%%%%%%%%%%%%%%%%%%%%%%%

\section{Electromagnetic transition} \label{sec:ffs}

\subsection{$N\Delta\gamma$ transition current}\label{sec:current}

             We now turn to the general properties of the $N\Delta\gamma$ transition current
             and its decomposition in terms of Lorentz-invariant form factors.
             The current can be generically written as
             \begin{equation}\label{current-general-0a}
                 J^{\mu,\rho}(P,Q) = \mathds{P}^{\rho\alpha}(P_f) \,i\gamma_5 \, \Gamma^{\alpha\mu}(P,Q) \,\Lambda_+(P_i) \,,
             \end{equation}
             where $P_i$ and $P_f$ are the incoming nucleon and outgoing $\Delta$ momenta, with $P_i^2=-M_N^2$ and $P_f^2=-M_\Delta^2$.
             They can be expressed through the photon momentum $Q=P_f-P_i$ and the average momentum $P=(P_i+P_f)/2$.
             The onshell structure of the current is ensured by the projectors defined in Eq.~\eqref{projectors}, i.e.,
             the positive-energy projector $\Lambda_+$ for the nucleon and the Rarita-Schwinger projector $\mathds{P}^{\rho\alpha}$ for the $\Delta$-baryon.
             Eq.~\eqref{current-general-0a} is a matrix in spinor space; the usual current matrix element $\langle P_f, s_f \,|\, J^\mu\,| \,P_i, s_i \rangle$ is obtained
             upon contraction with the $\Delta$ and nucleon spinors from Eq.~\eqref{spinors}.
             The momentum dependence of the projectors implies that the $\gamma-$matrices contained in the Rarita-Schwinger projector $\mathds{P}^{\rho\alpha}(P_f)$
             are now transverse with respect to $P_f$.
             We extracted an explicit factor $\gamma_5$ in Eq.~\eqref{current-general-0a} so that the remainder $\Gamma^{\alpha\mu}$, which will be specified below, has positive parity.

             Similarly to the nucleon and $\Delta$ bound-state amplitudes,
             the composition of the four-point function $\Gamma^{\alpha\mu}$ in Eq.~\eqref{current-general-0a} is determined by Poincar\'e covariance.
             For its explicit construction it is again convenient to work with orthogonal momenta. This is not yet the case for $P$ and $Q$ because
             the non-vanishing $N$-$\Delta$ mass difference
             entails \mbox{$P\cdot Q \neq 0$}, cf.~Eq.~\eqref{PdotQ}.
             We take instead the component of $P$ transverse to $Q$:
             \begin{equation}\label{PT}
                  P_T^\mu = T^{\mu\nu}_Q P^\nu = P^\mu - (P\cdot \widehat{Q})\,\widehat{Q}^\mu\,,
             \end{equation}
             and normalize it to unity: $K^\mu := \widehat{P_T}^\mu$. Here,
             \begin{equation}
                 T^{\mu\nu}_Q=\delta^{\mu\nu}-\widehat{Q}^\mu \widehat{Q}^\nu
             \end{equation}
             is the transverse projector with respect to $Q$. Together with the normalized photon momentum $\widehat{Q}$,
             the current is now characterized by two orthonormal four-momenta, $K$ and $\widehat{Q}$
             (instead of $P$ and $Q$, or $P_i$ and $P_f$), which will simplify its structure considerably.

             Using this construction, the most general form of the vertex $\Gamma^{\alpha\mu}$ that is
             compatible with Poincar\'e covariance, positive parity and current conservation
             can be written as (cf.~App.~\ref{sec:currentgeneral}):
             \begin{equation}\label{current-general1}
                 \Gamma^{\alpha\mu} = i \widehat{Q}^\alpha \left( g_1 \gamma^\mu_T + g_2 \,K^\mu\right) - g_3\,T_Q^{\alpha\mu} \,,
             \end{equation}
             where $\gamma^\mu_T$ is transverse to $Q$. It depends on three real and dimensionless form factors $g_i(Q^2)$.

             For comparison with experiment, it is more convenient to work with the Jones-Scadron form factors
             $G_M^\star(Q^2)$, $G_E^\star(Q^2)$ and $G_C^\star(Q^2)$ which are related to the
             pion electroproduction multipole amplitudes at the $\Delta-$resonance position % [PV, Eq. 2.13],
             and can be expressed in terms of helicity amplitudes~\cite{Jones:1972ky,Pascalutsa:2006up}.
             The respective decomposition of the vertex $\Gamma^{\alpha\mu}$ is:
             \begin{equation}\label{current-general-0b}
             \begin{split}
                 \Gamma^{\alpha\mu}  &= b\,\bigg[ \frac{i\omega}{2\lambda_+}\,(G_M^\star-G_E^\star)\,\gamma_5 \,\varepsilon^{\alpha\mu\gamma\delta} K^\gamma \widehat{Q}^\delta \\
                 &\qquad\qquad - G_E^\star \, T^{\alpha\gamma}_Q \,T^{\gamma\mu}_K - \frac{i\tau}{\omega}\,G_C^\star\,\widehat{Q}^\alpha K^\mu \bigg]\,,
             \end{split}
             \end{equation}
             where we used the dimensionless variables
             \begin{equation}\label{tau-lambda}
                 \tau := \frac{Q^2}{2\,(M_\Delta^2+M_N^2)}  \,, \;\;
                 \lambda_\pm := \frac{(M_\Delta \pm M_N)^2 + Q^2}{2\,(M_\Delta^2+M_N^2)}
             \end{equation}
             as well as $\omega:= \sqrt{\lambda_+ \lambda_-}$ and $b :=\sqrt{\tfrac{3}{2}} \,(1 + M_\Delta/M_N)$.
             We show in App.~\ref{sec:currentgeneral} that the vertices in~\eqref{current-general1} and~\eqref{current-general-0b} are equivalent
             when contracted with the projectors in the current matrix~\eqref{current-general-0a},
             and the relations between the $g_i$ and the Jones-Scadron form factors are stated in Eq.~\eqref{g-vs-JonesScadron}.
%             see also App.~\ref{app:kinematics}.

             Eq.~\eqref{current-general-0b} is identical with the standard
             Jones-Scadron expression~\cite{Jones:1972ky,Pascalutsa:2006up}
             which is given in terms of the Lorentz structures
             \begin{equation}\label{ndg-traditional}
             \begin{split}
                 \frac{\varepsilon^{\alpha\mu\gamma\delta} P_i^\gamma P_f^\delta}{M_\Delta^2+M_N^2}
                            &=  i\omega \, \varepsilon^{\alpha\mu\gamma\delta} K^\gamma \widehat{Q}^\delta \,, \\
                 \frac{\varepsilon^{\alpha\lambda\gamma\delta} P_i^\gamma P_f^\delta \,\varepsilon^{\mu\lambda\rho\sigma} P_i^\rho P_f^\sigma}{(M_\Delta^2+M_N^2)^2}
                            &= -\omega^2 \,T^{\alpha\gamma}_Q \,T^{\gamma\mu}_K\,, \\
                 \frac{Q^\alpha \left( Q^2 P^\mu - P\cdot Q \,Q^\mu \right)}{(M_\Delta^2+M_N^2)^2}
                            &= 2i  \omega \tau \,\widehat{Q}^\alpha K^\mu\,.
             \end{split}
             \end{equation}
             These relations can be verified by expressing $P_i$ and $P_f$ through $P$ and $Q$
             and subsequently in terms of the unit vectors $K$ and $\widehat{Q}$ via Eq.~\eqref{PQ-vs-KQHat}.

         %!!!!!!!!!!!!!!!!!!!!!!!!!!!!!!!!!!!!!!!!!!!!!!!!!!!!!!!!!!!!!!!!!!!!!!!!!!!!!!!!!!!!!!!!!!!!!!!!!!!!!!!!!!!!!!!

         \begin{figure}[t]
                    \begin{center}
                    \includegraphics[scale=0.11]{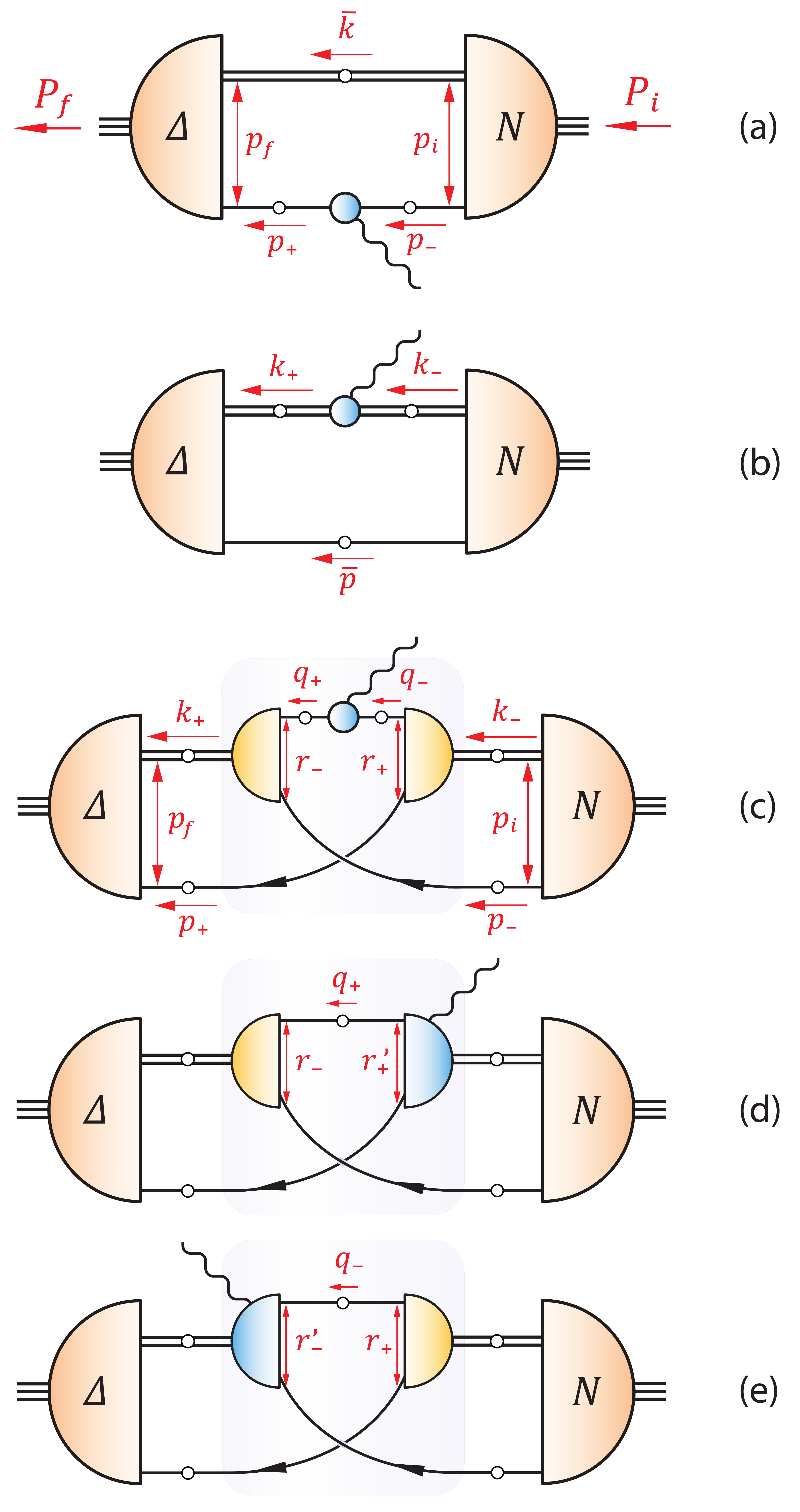}
                    \caption{(Color online) General expression for the $N\Delta\gamma$ transition current
                             in the quark-diquark approach, see Eqs.~\eqref{emcurrent-gauging} and App.~\ref{app:current-details}.  }\label{fig:current}
                    \end{center}
        \end{figure}

         %!!!!!!!!!!!!!!!!!!!!!!!!!!!!!!!!!!!!!!!!!!!!!!!!!!!!!!!!!!!!!!!!!!!!!!!!!!!!!!!!!!!!!!!!!!!!!!!!!!!!!!!!!!!!!!!

\subsection{Electromagnetic current in the quark-diquark approach}

         The computation of the $N\Delta\gamma$ transition matrix of
         Eqs.~\eqref{current-general-0a} and~\eqref{current-general-0b} from its substructure in QCD
         requires a microscopic description of its ingredients.
         A systematic construction principle to derive the coupling of a hadron to an external current
         is the 'gauging of equations' method of Refs.~\cite{Haberzettl:1997jg,Kvinikhidze:1998xn,Kvinikhidze:1999xp}.
         The procedure was applied in~\cite{Oettel:1999gc} to derive the relevant diagrams in the quark-diquark system;
         recent discussions and applications in the three-quark framework can be found in Refs.~\cite{Eichmann:2011vu,Eichmann:2011pv,Eichmann:2011ec}.

         Applied to our case, the basic idea is that the $N\Delta\gamma$ transition matrix element
         is the $N\Delta$ pole residue of the quark-diquark Green function that is struck by an external photon.
         If the current systematically couples to all internal constituents, which means that it has the formal properties of a derivative,
         electromagnetic current conservation is automatically satisfied.
         The photon coupling at the hadron level is thereby resolved to its interaction with quarks and diquarks.
         The resulting current has the generic form
             \begin{equation}\label{emcurrent-gauging}
                J^\mu  = \conjg{\Gamma}_\Delta \,G_0 \left(\mathbf{\Gamma}^\mu - K_\text{Q-DQ}^\mu \right) G_0 \,\Gamma_N \,,
             \end{equation}
         which is detailed in App.~\ref{app:current-details} and illustrated in Fig.~\ref{fig:current}.
         $\Gamma_N$ and $\Gamma_\Delta$ denote the incoming and outgoing baryon amplitudes, $G_0$
         is the disconnected product of a dressed quark and diquark propagator, and
         $\mathbf{\Gamma}^\mu$ represents the photon's coupling to $G_0$ which yields
         the impulse-approximation diagrams (a--b) in Fig.~\ref{fig:current}.
         Those alone are however not sufficient: in order to comply with
         electromagnetic gauge invariance, one must take the coupling to the quark-diquark kernel into account as well.
         This leads to an exchange-quark term (c) and further diagrams that involve seagulls (d--e), i.e.,
         the photon's coupling to the quark-diquark vertices.

         This decomposition has been frequently used in the computation of nucleon form factors by
         employing model ans\"atze for its ingredients~\cite{Oettel:2000jj,Cloet:2008re}.
         Building upon Refs.~\cite{Eichmann:2007nn,Eichmann:2008ef}, our goal in the present work is to compute these ingredients selfconsistently.
         We have outlined the calculation of quark and diquark propagators, diquark amplitudes,
         and nucleon and $\Delta$ bound-state amplitudes in the previous sections.
         In addition, we also compute the quark-photon and diquark-photon vertices that appear in diagrams (a--c),
         so that no further model input is needed for those objects.
         The only quantities that require a certain amount of modeling are the seagulls which we will discuss below.
         We note that the same method was recently applied for studying nucleon~\cite{Eichmann:2010je} and
         $\Delta$ form factors~\cite{Nicmorus:2010sd} in the quark-diquark approach and,
         upon neglecting the seagull terms, also the $\Delta N \pi$ transition form factor~\cite{Mader:2011zf}.

            The central ingredient of Fig.~\ref{fig:current} that describes the microscopic interaction with the photon
            is the dressed quark-photon vertex $\Gamma^\mu_\text{q}(k,Q)$.
            Ultimately, the diquark-photon and seagull terms would be resolved to the photon's coupling with quarks as well.
            The properties of the quark-photon vertex will therefore be reflected in all form-factor contributions,
            either directly via diagrams (a) and (c)  or implicitly in diagrams (b), (d) and (e).
            Here we compute the vertex self-consistently from its inhomogeneous Bethe-Salpeter equation~\cite{Maris:1999bh}:
             \begin{equation}\label{qpv-bse}
             \begin{split}
                  \Big[&\Gamma^\mu_\text{q}(k,Q)-Z_2\,i\gamma^\mu\Big]_{\alpha\beta}  = \\
                  & = \int\limits_{k'} \mathcal{K}_{\alpha\alpha'\beta'\beta}\,
                   \left[ S(k_+') \,\Gamma^\mu_\text{q}(k',Q)\,S(k_-')\right]_{\alpha'\beta'}\,,
             \end{split}
             \end{equation}
            see App.~D3 of Ref.~\cite{Eichmann:2011vu} for details of the solution method.
            The equation features the same rainbow-ladder kernel of Eq.~\eqref{RLkernel}
            that also appears in the quark DSE and diquark BSE.
            Electromagnetic gauge invariance is expressed by the vector Ward-Takahashi identity,
                  \begin{equation}\label{qpv-wti}
                      Q^\mu \,\Gamma^\mu_\text{q}(k,Q) = S^{-1}(k_+)-S^{-1}(k_-)\,,
                  \end{equation}
            which is satisfied for the resulting vertex. In combination with analyticity at $Q^2 \rightarrow 0$, it
            allows to write the vertex as the sum of a Ball-Chiu term~\cite{Ball:1980ay},
            which is fixed by gauge invariance, and a further transverse contribution:
                  \begin{equation}\label{vertex:BC}
                      \Gamma^\mu_\text{q}(k,Q) =   i\gamma^\mu\,\Sigma_A + 2 k^\mu (i\Slash{k}\, \Delta_A  + \Delta_B) + \Gamma^{\mu}_T\,,
                  \end{equation}
            where $A(k^2)$ and $B(k^2)$ are the dressing functions of the inverse quark propagator
            $S^{-1}(k)=i \Slash{k}\,A(k^2) + B(k^2)$, and we abbreviated
                 \begin{equation*}\label{QPV:sigma,delta}
                     \Sigma_F := \frac{F(k_+^2)+F(k_-^2)}{2} , \quad  \Delta_F := \frac{F(k_+^2)-F(k_-^2)}{k_+^2-k_-^2},
                  \end{equation*}
            with $k_\pm = k \pm Q/2$.
             As a consequence of the term $i\gamma^\mu$ that appears in Eq.~\eqref{qpv-bse},
             the structure of the BSE implies the existence of a transverse vector-meson pole in the quark-photon vertex~\cite{Maris:1999bh}, see~\cite{Eichmann:2011pv} for a recent discussion.
             Indeed, the inhomogeneous BSE self-consistently generates a $\rho-$meson pole in the vertex at $Q^2 =-m_\rho^2$
             which dominates the timelike structure of the $N\Delta\gamma$ transition form factors.

             Having determined the quark-photon vertex, the diquark-photon vertices that appear in Fig.~(\ref{fig:current}b) are computed selfconsistently as well.
             The diquark content of the nucleon and $\Delta$ amplitudes necessitates the inclusion of axial-axial as well as axial-scalar transition vertices.
             Applying the gauging procedure to the inverse diquark propagator of Eq.~\eqref{diquark-propagator} resolves the diquark-photon vertices
             into their quark-gluon substructure, where the photon couples once again to all constituents and thereby ensures
             a Ward-Takahashi identity for the vertex~\cite{Kvinikhidze:1999xp,Oettel:2002wf}.
             This leads to various quark-loop diagrams that involve the quark-photon vertex, seagull terms,
             and also the $qq$ kernel. The construction is detailed in App.~A.7 of Ref.~\cite{Eichmann:2009zx} and can be directly adopted here.
             The properties of the quark-photon vertex and, in particular, the $\rho-$meson pole that appears in its
             transverse term, are thereby recovered in the diquark-photon vertices as well.

            Finally, current conservation in the quark-diquark approach also requires the inclusion of seagull vertices $M^{\mu,\alpha}$ in diagrams~(\ref{fig:current}d)--(\ref{fig:current}e).
            These terms describe the coupling of the photon to the offshell quark-diquark vertices.
            As they are another source of model uncertainty they warrant further discussion.
            The seagull vertices $M^{\mu,\alpha}$ satisfy a Ward-Takahashi identity similar to Eq.~\eqref{qpv-wti}
            which involves differences of scalar and axial-vector diquark amplitudes~\cite{Wang:1996zu,Oettel:1999gc,Eichmann:2007nn}.
            In analogy to the Ball-Chiu construction for the quark-photon vertex,
            the seagull vertex can be written as the sum of a part that is fixed by the WTI and analyticity at $Q^2\rightarrow 0$,
            augmented by a further unconstrained transverse piece:
            \begin{equation}\label{ff:seagull}
                M^{\mu,\alpha} = M^{\mu,\alpha}_\text{WTI} + M^{\mu,\alpha}_T \,.
            \end{equation}
            The first term $M^{\mu,\alpha}_\text{WTI}$ is a rather lengthy expression,
            especially once the full Dirac-Lorentz substructure of the diquarks is taken into
            account. Its detailed form is given in App.~A.8 of Ref.~\cite{Eichmann:2009zx}; here we only repeat the generic structure:
            \begin{equation}\label{seagulls-generic}
            \begin{split}
                M_\text{WTI} &= e_- M_1 + e_+ M_2 -e_{dq} M_3\,, \\
                \conjg{M}_\text{WTI} &= -\tilde{e}_+ \conjg{M}_1 - \tilde{e}_- \conjg{M}_2 + \tilde{e}_{dq} \conjg{M}_3\,.
            \end{split}
            \end{equation}
            The first and second lines correspond to the expressions that enter Figs.~(\ref{fig:current}d) and~(\ref{fig:current}e), respectively.
            The flavor-charge traces $e_\pm$ and $e_{dq}$ represent the photon coupling to the seagulls' quark and diquark legs
            and are determined in App.~\ref{app:color-flavor}.

            Since the properties of the quark-photon vertex in principle also dictate the structure of the seagulls,
            the transverse term in~\eqref{ff:seagull} must involve a $\rho-$meson pole as well.
            In absence of a selfconsistent solution for the seagulls, we use the following ansatz for the $\rho-$meson part:
            \begin{equation}\label{ff:seagull-rho}
                M^{\mu,\alpha}_T = -g(x)\,T_Q^{\mu\nu}\,M^{\nu,\alpha}_\text{WTI}  \,,
            \end{equation}
            where $x=Q^2/m_\rho^2$.
            Purely longitudinal terms in~\eqref{ff:seagull} do not contribute to the $N\Delta\gamma$ current because of current conservation;
            hence, this simple form amounts to an overall factor $(1-g(x))$ in the WTI-conserving part.
            The following phenomenological function was chosen in Ref.~\cite{Eichmann:2008ef}:
            \begin{equation}\label{ff:seagull-rho-2}
                g(x) = \frac{1}{g_\rho}\,\frac{x^2}{1+x}\,e^{-\rho_3 (1+x)} \,,
            \end{equation}
            where $g_\rho = \sqrt{2}\,m_\rho/f_\rho$ is computed from the $\rho-$meson BSE. The parameter $\rho_3$ will be discussed below;
            it contributes to the overall model uncertainty that is, so far, induced by the width parameter $\eta$ in the effective coupling~\eqref{couplingMT} only.

               %!!!!!!!!!!!!!!!!!!!!!!!!!!!!!!!!!!!!!!!!!!!!!!!!!!!!!!!!!!!!!!!!!!!!!!!!!!!!!!!!!!!!!!!!!!!!!!!!!!!!!!!!!!!!!!!

               \begin{figure}[t]
                          \begin{center}

                          \includegraphics[scale=1.15]{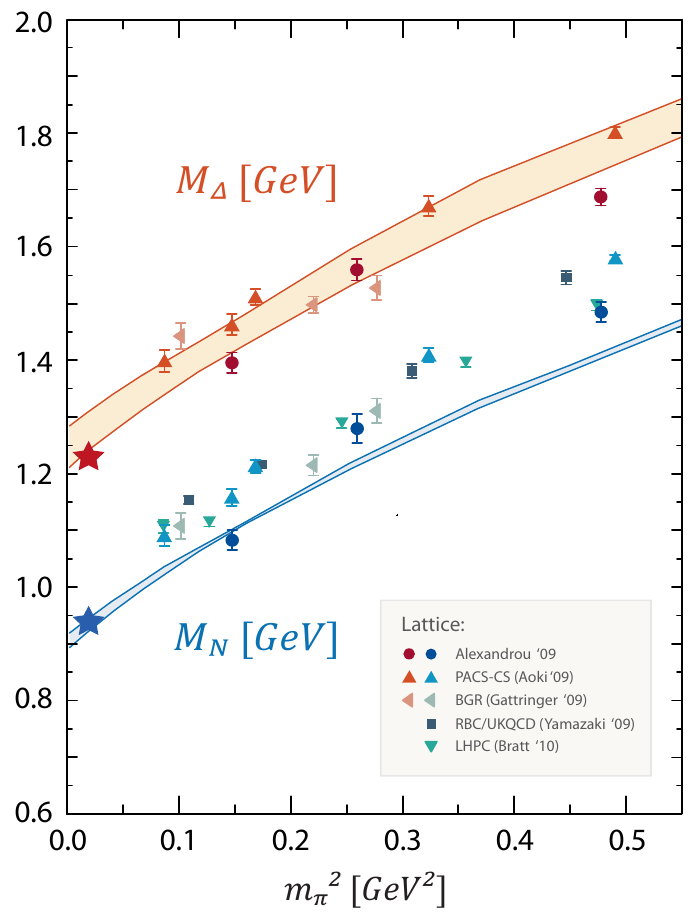}
                          \caption{(Color online) Current-mass evolution of nucleon and $\Delta$ masses from their quark-diquark BSEs.
                                   The band corresponds to a variation of the width parameter $\eta=1.8\pm 0.2$;
                                   the star denotes the experimental value.
                                   We compare to a selection of lattice data for nucleon~\cite{Alexandrou:2009hs,Aoki:2008sm,Gattringer:2008vj,Yamazaki:2009zq,Bratt:2010jn}
                                   and $\Delta$~\cite{Alexandrou:2009hs,Aoki:2008sm,Gattringer:2008vj}.}\label{fig:nucleonmass}

                          \end{center}
              \end{figure}

               %!!!!!!!!!!!!!!!!!!!!!!!!!!!!!!!!!!!!!!!!!!!!!!!!!!!!!!!!!!!!!!!!!!!!!!!!!!!!!!!!!!!!!!!!!!!!!!!!!!!!!!!!!!!!!!!

%%%%%%%%%%%%%%%%%%%%%%%%%%%%%%%%%%%%%%%%%%%%%%%%%%%%%%%%%%%%%%%%%%%%%%%%%%%%%%%%%%%%%%%%%%%%%%%%%%%%%%%%%%%%%%%%%%%%%%%%%%%%

\section{Results and discussion} \label{sec:results}

          We summarize the steps that we performed to obtain the results in this section.
          The basic model input is the effective quark-gluon interaction of Eq.~\eqref{couplingMT} that enters the rainbow-ladder kernel of Eqs.~\eqref{RLkernel}.
          Its infrared part depends on the scale $\Lambda=0.72$~GeV which we adjust to reproduce the experimental pion decay constant, whereas
          the width $\eta=1.8 \pm 0.2$ remains a parameter and modifies the infrared shape of the coupling $\alpha(k^2)$.
          Using this input, we solve the Dyson-Schwinger equation~\eqref{quarkdse} for the quark propagator,
          the diquark Bethe-Salpeter equation~\eqref{diquark-bse} for the scalar and axialvector diquark amplitudes,
          and we compute the diquark propagators from Eq.~\eqref{diquark-propagator}.
          These quantities are subsequently implemented in the quark-diquark BSEs~\eqref{quark-diquark-bse} from which we obtain
          the nucleon and $\Delta$ masses and bound-state amplitudes.

          The resulting masses for nucleon and $\Delta$ were obtained in Refs.~\cite{Eichmann:2008ef,Nicmorus:2010sd,Nicmorus:2010mc}
          and are shown in Fig.~\ref{fig:nucleonmass} as a function of the squared pion mass.
          They are in reasonable agreement with experiment and lattice results.
          The results obtained in the quark-diquark model are in several respects similar
          to recent three-body calculations of nucleon and $\Delta$ properties~\cite{Eichmann:2009qa,SanchisAlepuz:2011jn,Eichmann:2011vu}.
          The nucleon mass is practically identical in both approaches, whereas the $\Delta$ mass in the quark-diquark model
          is larger by $\lesssim 5\%$ and also shows a larger dependence on $\eta$ which is illustrated by the bands.
          Analogous observations also hold for the nucleon's electromagnetic form factors which are quantitatively similar
          in both setups except that the quark-diquark approach produces a larger model dependence.
          Note that the rainbow-ladder truncation does not dynamically generate a $\Delta\rightarrow N\pi$ decay width, i.e.,
          our result for the $\Delta-$baryon describes a stable bound state.
          Associated non-analyticities which would appear for $M_\Delta-M_N>m_\pi$ are therefore absent.

\begin{figure}[t]
                           \begin{center}
                           \includegraphics[scale=0.36]{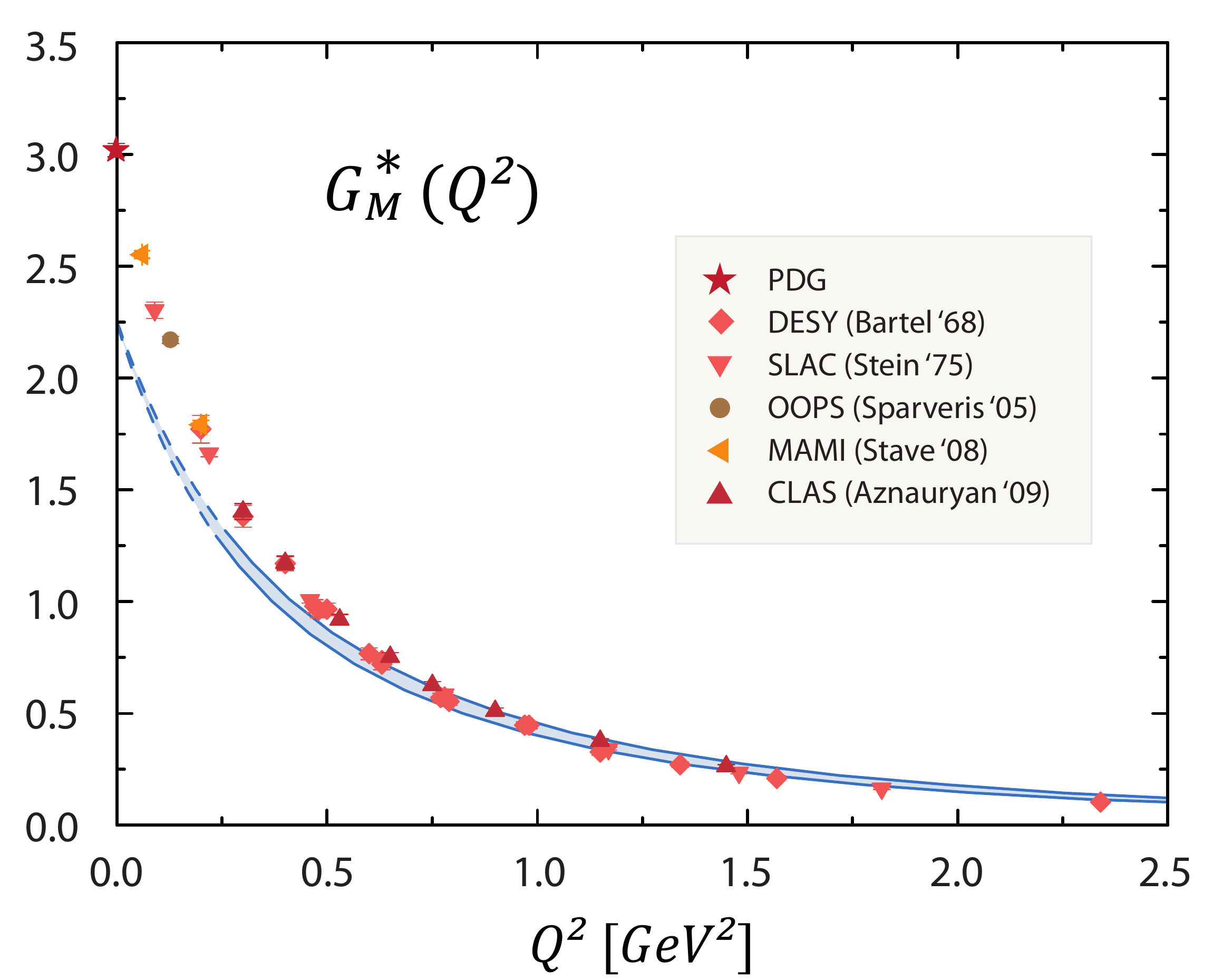}
                           \caption{(Color online) $Q^2-$evolution of the magnetic dipole form factor $G_M^\star$ in comparison with
                                    experimental data from Refs.~\cite{Bartel:1968tw,Stein:1975yy,Sparveris:2004jn,:2008tv,Aznauryan:2009mx}.
                                    The band denotes the model dependence as discussed in the text. }\label{fig:GM}
                           \end{center}
               \end{figure}

\begin{figure*}[t]
                           \begin{center}
                           \includegraphics[scale=0.36]{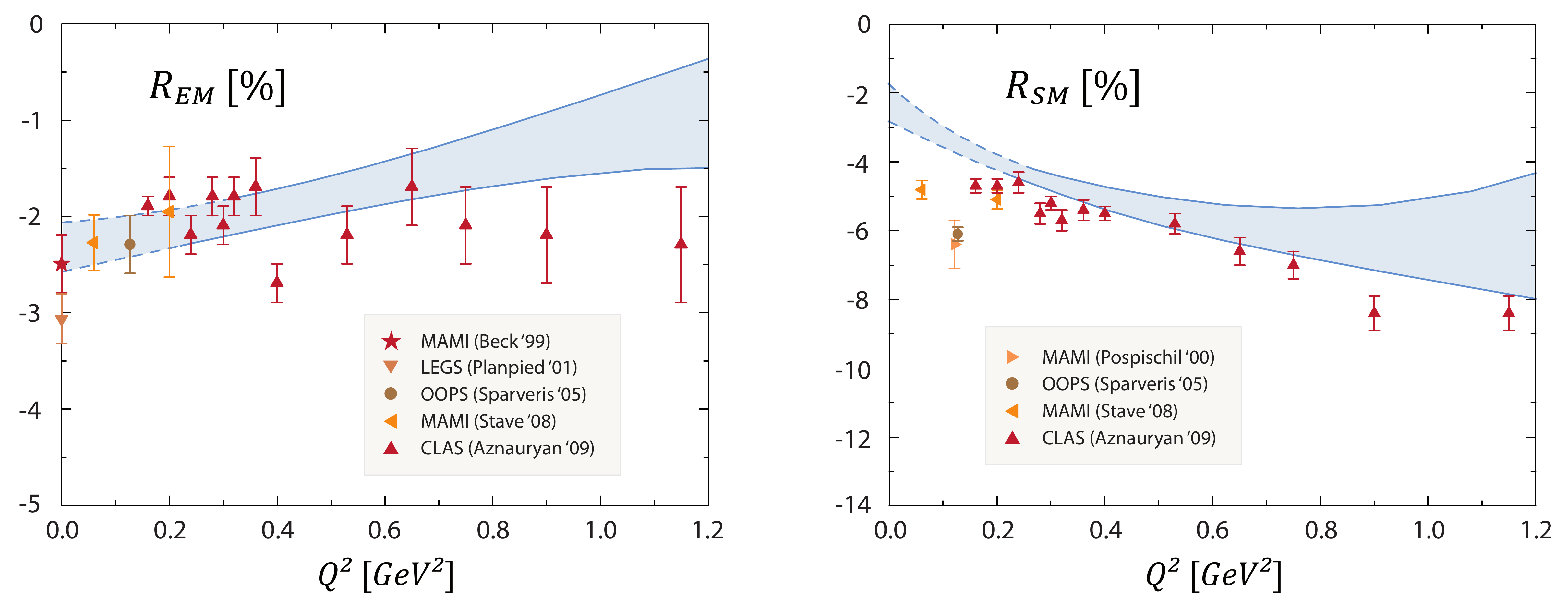}
                           \caption{(Color online) Form-factor ratios $R_{EM}(Q^2)$ and $R_{SM}(Q^2)$ compared to
                                     experimental data from Refs.~\cite{Beck:1999ge,Pospischil:2000ad,Blanpied:2001ae,Sparveris:2004jn,:2008tv,Aznauryan:2009mx}.
                                     The bands denote the model uncertainty as described in the text.  }\label{fig:REM-RSM}
                           \end{center}
               \end{figure*}

          Having established the ingredients of the electromagnetic transition current, we calculate its matrix elements
          from the diagrams in Fig.~\ref{fig:current} which are discussed in detail in App.~\ref{app:current-details}.
          The current implements the selfconsistent solution for the quark-photon vertex~\eqref{qpv-bse} and the diquark-photon vertex,
          and for the seagull vertices we use the expression in Eq.~\eqref{ff:seagull}.
          The seagulls, which are necessary for current conservation in the quark-diquark model, represent another source of model uncertainty.
          Their transverse $\rho-$meson parts are modeled by Eqs.~(\ref{ff:seagull-rho}--\ref{ff:seagull-rho-2}) and affect the larger-$Q^2$ behavior of the form factors.
          We take a variation of the parameter $\rho_3 \in [ 0, 0.15]$ into account;
          the central value of that interval was used in Ref.~\cite{Eichmann:2008ef} to maximize agreement for nucleon electromagnetic form factors at larger $Q^2$.
          The combined model dependence, stemming from the the seagull variation together with the $\eta$ dependence in the effective interaction,
          leads to the colored bands in Figs.~(\ref{fig:GM}--\ref{fig:GE-partialwaves}).

          The Jones-Scadron form factors $G_M^\star(Q^2)$, $G_E^\star(Q^2)$ and $G_C^\star(Q^2)$ are finally extracted from the Dirac traces in Eq.~\eqref{current-traces}.
          Since the approach is Poincar\'e-covariant, the results are independent of the choice of reference frame.
          In order to avoid complex continuations for the radial momentum variables in the $N$ and $\Delta$ bound-state amplitudes,
          we work in the frame where the photon momentum is purely real: $Q=(0,0,|Q|,0)$ or, expressed in terms
          of the unit vectors defined in Section~\ref{sec:current}, $\widehat{Q}=e_3$ and $K=e_4$. The singularities in the quark and diquark propagators
          that enter the form factor integrals restrict the accessible domain of photon momenta to $Q^2 \lesssim 2.5$~GeV$^2$, see App.~\ref{app:kinematic-restrictions}.
          This value is quite small and due to the quark-diquark description; a genuine three-body calculation would be able to reach $Q^2$ values roughly twice as large.
          In addition, the kinematic dependence on the non-vanishing $N$-$\Delta$ mass difference also imposes  a lower limit for $Q^2$.
          In order to obtain results at $Q^2=0$, we extrapolate the form factor results at non-zero momentum transfer using Pad\'e approximants.
          The extrapolation regions are indicated by the dashed margins in Figs.~(\ref{fig:GM}--\ref{fig:REM-RSM}).

               %!!!!!!!!!!!!!!!!!!!!!!!!!!!!!!!!!!!!!!!!!!!!!!!!!!!!!!!!!!!!!!!!!!!!!!!!!!!!!!!!!!!!!!!!!!!!!!!!!!!!!!!!!!!!!!!

               %!!!!!!!!!!!!!!!!!!!!!!!!!!!!!!!!!!!!!!!!!!!!!!!!!!!!!!!!!!!!!!!!!!!!!!!!!!!!!!!!!!!!!!!!!!!!!!!!!!!!!!!!!!!!!!!

               %!!!!!!!!!!!!!!!!!!!!!!!!!!!!!!!!!!!!!!!!!!!!!!!!!!!!!!!!!!!!!!!!!!!!!!!!!!!!!!!!!!!!!!!!!!!!!!!!!!!!!!!!!!!!!!!

                \renewcommand{\arraystretch}{1.0}

                \begin{table}[t]

                     \begin{center}

                     \begin{tabular}{    @{\;\;}c@{\;\;} || @{\;\;}c@{\;\;} | @{\;\;}c@{\;\;} ||   @{\;\;}c@{\;\;} | @{\;\;}c@{\;\;} | @{\;\;}c@{\;\;}     }

                                &  $M_N$      &  $M_\Delta$    &  $G_M^\star(0)$  &  $R_{EM}(0)$  &  $R_{SM}(0)$       \\   \hline

                         Exp.   &  $0.94$     &  $1.23$        &  $3.02(3)$       &  $-2.5(5)$    &                            \\
                         Calc.  &  $0.94(1)$  &  $1.27(3)$     &  $2.23(2)$       &  $-2.3(3)$    &  $-2.2(6)$

                     \end{tabular}

                     \caption{Results at the physical $u/d$ mass compared to experiment. Nucleon and $\Delta$ masses are in units of GeV,
                              $G_M^\star(0)$ is dimensionless, and the ratios $R_{EM}$ and $R_{SM}$ are given in percent.
                              The experimental values for $G_M^\star(0)$ and $R_{EM}(0)$ are the PDG values~\cite{Nakamura:2010zzi}.
                              The parentheses in our results indicate the combined model dependence as discussed in the text.
                                            }\label{tab:results}
                     \end{center}

                \end{table}

               %!!!!!!!!!!!!!!!!!!!!!!!!!!!!!!!!!!!!!!!!!!!!!!!!!!!!!!!!!!!!!!!!!!!!!!!!!!!!!!!!!!!!!!!!!!!!!!!!!!!!!!!!!!!!!!!

               %!!!!!!!!!!!!!!!!!!!!!!!!!!!!!!!!!!!!!!!!!!!!!!!!!!!!!!!!!!!!!!!!!!!!!!!!!!!!!!!!!!!!!!!!!!!!!!!!!!!!!!!!!!!!!!!

               %!!!!!!!!!!!!!!!!!!!!!!!!!!!!!!!!!!!!!!!!!!!!!!!!!!!!!!!!!!!!!!!!!!!!!!!!!!!!!!!!!!!!!!!!!!!!!!!!!!!!!!!!!!!!!!!

          \subsection{$Q^2$ dependence of the form factors} \label{sec:results-Q2}

             The $N\Delta\gamma$ transition current is determined by the three Jones-Scadron form factors $G_M^\star(Q^2)$, $G_E^\star(Q^2)$ and $G_C^\star(Q^2)$
             which are experimentally extracted from the multipole amplitudes in pion electroproduction~\cite{Pascalutsa:2006up,Aznauryan:2011qj}.
             The process is dominated by a magnetic dipole transition $(M1)$ which,
             in a quark-model picture, amounts to a spinflip of a quark and is encoded in the form factor $G_M^\star(Q^2)$.
             Its static experimental value is $G_M^\star(0)=3.02(3)$~\cite{Nakamura:2010zzi};
             experimental data exist in the range up to $Q^2\sim 8$~GeV$^2$.
             The remaining electric $(E2)$ and Coulomb $(C2)$ quadrupole contributions are much smaller and measure the deformation in the transition.
             They are expressed by the form factors $G_E^\star(Q^2)$ and $G_C^\star(Q^2)$ which are
             usually related to the magnetic dipole form factor through the form factor ratios
             \begin{equation}\label{ratios}
                 R_{EM} = -\frac{G_E^\star}{G_M^\star}\,, \quad
                 R_{SM} = -\frac{|\vect{Q}|}{2M_\Delta}\,\frac{G_C^\star}{G_M^\star}\,,
             \end{equation}
             where $|\vect{Q}|$ denotes the magnitude of the photon three-momentum in the $\Delta$ rest frame.
             It can be expressed in terms of Lorentz-invariant variables via
             \begin{equation}
                 \frac{|\vect{Q}|}{2M_\Delta} = \frac{\omega}{1+2\delta}\,,
             \end{equation}
             where $\omega$ was defined below Eq.~\eqref{tau-lambda} and $\delta$ is related to the $N$--$\Delta$ mass difference, cf.~Eq.~\eqref{kinematics-variables}:
             \begin{equation}
                 \delta = \frac{M_\Delta^2-M_N^2}{2\,(M_\Delta^2+M_N^2)}\,.
             \end{equation}

             Our result for the magnetic dipole form factor $G_M^\star(Q^2)$ is shown in Fig.~\ref{fig:GM}.
             We find good agreement with experimental data above $Q^2\sim 1$~GeV$^2$, whereas the quark-diquark result underestimates these
             data by $\sim 25\%$ in the limit $Q^2=0$, cf.~Table~\ref{tab:results}.
             This is comparable to constituent-quark model predictions~\cite{Faessler:2006ky}, where the
             long-standing discrepancy with the data has been attributed to missing meson-cloud contributions.
             Their impact has been studied with dynamical reaction models~\cite{Kamalov:2000en,Sato:2000jf},
             where the 'bare' $\Delta$ resonance extracted from the $N\gamma^\star\to N\pi$ scattering amplitude
             corresponds to the quark-core contribution and meson-cloud effects
             are generated via rescattering processes.  % $t-$channel meson exchange, followed by
             In these analyses the pion cloud is sizeable and accounts for $\sim 30\%$ of $G_M^\star(0)$.
             Similar conclusions have been found in the cloudy bag model~\cite{Bermuth:1988ms,Lu:1996rj}    % by including pionic components, e.g.,
             or covariant chiral quark models~\cite{Faessler:2006ky}.

             The same interpretation applies to our framework as well.
             Pion-loop effects are not implemented in a rainbow-ladder truncation,
             and their inclusion would yield characteristic non-analytic structures in the chiral and low-momentum structure of form factors.
             Our result for $G_M^\star(Q^2)$ is consistent with various nucleon form-factor calculations in the Dyson-Schwinger/Faddeev approach,
             where similar discrepancies were interpreted as signals of missing pion-cloud effects~\cite{Eichmann:2010je,Eichmann:2011vu}.
             Typical examples are the nucleon's charge radii which underestimate the experimental values but converge
             with lattice data at larger quark masses; or the low-$Q^2$ behavior of electromagnetic form factors that shows missing structure,
             whereas one finds reasonable agreement with experiment at larger $Q^2$.
             A recent calculation of the nucleon's axial charge finds such a discrepancy as well~\cite{Eichmann:2011pv}.
             On the other hand, the nucleon's isoscalar anomalous magnetic moment $\kappa_s$, where leading-order chiral corrections cancel,
             is accurately reproduced by the Faddeev calculation.
             These observations suggest to identify the rainbow-ladder truncated nucleon with the 'quark core' in chiral effective field theories.

               %!!!!!!!!!!!!!!!!!!!!!!!!!!!!!!!!!!!!!!!!!!!!!!!!!!!!!!!!!!!!!!!!!!!!!!!!!!!!!!!!!!!!!!!!!!!!!!!!!!!!!!!!!!!!!!!

                \begin{figure*}[t]
                           \begin{center}
                           \includegraphics[scale=0.36]{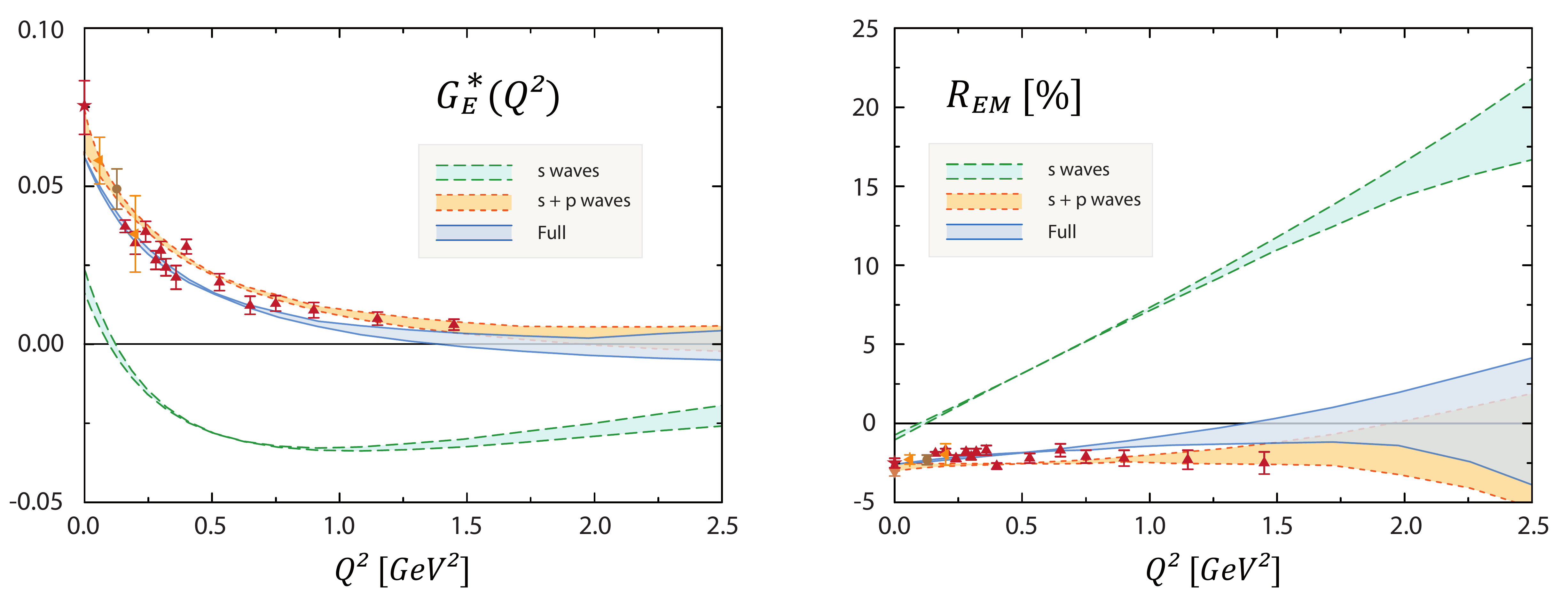}
                           \caption{(Color online) Decomposition of $G_E^\star(Q^2)$ and $R_{EM}(Q^2)$ according to the orbital angular-momentum content of nucleon and $\Delta$ amplitudes.
                                    The dashed bands are the results obtained with the $s-$wave basis elements in Table~\ref{fig:partial-wave}
                                    and the dotted bands follow upon including the $p-$wave elements as well.
                                    The experimental data are the same as in Fig.~\ref{fig:REM-RSM}.   }\label{fig:GE-partialwaves}
                           \end{center}
               \end{figure*}

               %!!!!!!!!!!!!!!!!!!!!!!!!!!!!!!!!!!!!!!!!!!!!!!!!!!!!!!!!!!!!!!!!!!!!!!!!!!!!!!!!!!!!!!!!!!!!!!!!!!!!!!!!!!!!!!!

             In fact, we observe analogous discrepancies in the low-$Q^2$ region for all three form factors $G_M^\star(Q^2)$, $G_E^\star(Q^2)$ and $G_C^\star(Q^2)$.
             Consequently, these deviations cancel in the ratios of Eq.~\eqref{ratios}, and $R_{EM}(Q^2)$ and $R_{SM}(Q^2)$
             show a good agreement with the experimental data which can be seen in Fig.~\ref{fig:REM-RSM}.
             Considering the absence of pion-cloud effects in our description, this result is quite remarkable.
             Non-zero values for $R_{EM}(0)$ and $R_{SM}(0)$ are usually attributed to the presence of
             quark orbital angular momentum in the $N\Delta\gamma$ transition,
             either via $d$-wave components in the nucleon and $\Delta$ wave functions, or caused by pion-cloud contributions.
             A large sensitivity to pionic effects has been found in coupled-channel analyses as well, where the bare ratios
             extracted from the electroproduction data are close to zero, so that $R_{EM}$ and $R_{SM}$ would be almost
             entirely dominated by the meson cloud~\cite{Kamalov:2000en,Sato:2000jf}.

             A related question concerns the asymptotic behavior of the form factors $R_{EM}$ and $R_{SM}$
             and the scale where a perturbative-QCD description sets in.
             Dimensional counting rules and hadron helicity conservation predict~\cite{Carlson:1985mm}
             \begin{equation}\label{ratios-perturbative}
                 R_{EM} \to 1\,, \quad
                 R_{SM} \to const. \quad
                 \text{for} \quad Q^2\to\infty\,,
             \end{equation}
             which is clearly not realized for the available experimental data. $R_{EM}$ is negative and small
             and remains practically constant in the entire $Q^2$ range whereas $R_{SM}$ is negative but rises in magnitude at larger $Q^2$.
             Taking into account the orbital motion of the partons in the nucleon and $\Delta$ wave functions leads to a
             double-logarithmic correction in the ratio $R_{SM}$ of Eq.~\eqref{ratios-perturbative}~\cite{Idilbi:2003wj}.

              While the large-$Q^2$ region remains inaccessible to our analysis due to the kinematic singularity restrictions,
              we can investigate the impact of quark orbital angular-momentum correlations in the bound-state amplitudes.
              As a consequence of Poincar\'e covariance, the nucleon and $\Delta$ wave functions do not only include $s$ and $d-$wave components
              but also $p$ waves and, in the case of the $\Delta$, even $f$ waves.
              The corresponding basis decomposition in the quark-diquark model,
              in terms of eigenvalues of quark-diquark spin and orbital angular momentum in the respective rest frames,
              is shown in Table~\ref{fig:partial-wave}.
              While the interpretation of spin and orbital angular momentum can change in different frames,
              the decomposition itself is Lorentz-covariant, and basis elements which correspond to $p$ waves in the rest frame
              carry one power of the relative quark-diquark momentum unit $r^\mu$.
              Our following use of the terminology '$s$ waves' and '$p$ waves' will therefore refer to that basis decomposition.

              The analysis of the $N$ and $\Delta$ dressing functions associated with the dimensionless basis elements in Table~\ref{fig:partial-wave} exhibits
              a clear hierarchy in their magnitude: $s$ waves are dominant, $p$ waves are suppressed, and $d$ waves provide a tiny contribution.
              The same observation can be made in the three-body description of the nucleon amplitude, where
              $s$ waves contribute roughly $\nicefrac{2}{3}$ to the nucleon's normalization and $p$ waves the remaining third,
              whereas the contribution from $d$ waves is at the order of only one percent~\cite{Eichmann:2011vu}.
              Thus, one would expect the dominant orbital angular-momentum effects in the form factors to come from $p$-wave contributions as well.

              In order to test the sensitivity of the $N\Delta\gamma$ transition to different orbital angular-momentum correlations,
              we computed the Jones-Scadron form factors upon retaining $s-$wave elements only.
              This leads to a notably different behavior for the  electric quadrupole transition $G_E^\star(Q^2)$, shown in Fig.~\ref{fig:GE-partialwaves}.
              Except for very small $Q^2$, its result carries now a negative sign.
              The corresponding value for the ratio $R_{EM}$ starts off close to zero and rises almost linearly with increasing photon momentum.
              Although the upper $Q^2$ limit is certainly too small to allow any statements concerning the perturbative behavior,
              a trend towards the perturbative prediction $R_{EM}\to 100\%$ in Eq.~\eqref{ratios-perturbative} is visible.
              If $p$ waves are included, $R_{EM}$ changes its behavior and becomes negative, thereby reinstating agreement with the experimental data.
              The inclusion of further $d-$ and $f-$wave basis elements produces only a minor change in that result.
              This observation implies that the negative value for $R_{EM}$ is indeed an effect of quark orbital angular momentum;
              however, it predominantly owes to $p-$wave effects in the amplitudes which
              are already generated from the dynamics of the quark core as a consequence of Poincar\'e covariance.
              This demonstrates the importance of relativistic effects in the properties of the $N\Delta\gamma$ transition.
              Such effects are missed in the non-relativistic quark model, or in a covariant description where $p$ waves are not taken into account.

              On the other hand, we find no such behavior for the remaining form factors $G_M^\star(Q^2)$ and $G_C^\star(Q^2)$.
              The result for $G_M^\star(Q^2)$ obtained with $s$ waves only is practically indistinguishable from Fig.~\ref{fig:GM} throughout the $Q^2$ range,
              which indicates that orbital effects do no play a role in the magnetic transition form factor.
              The same is true for the Coulomb ratio $R_{SM}$ which exhibits a similar shape as the full result in Fig.~\ref{fig:REM-RSM}, except for a broader
              model uncertainty.

              We finally comment on the importance of axialvector-diquark degrees of freedom in the nucleon amplitude.
              Due to its isospin-$\nicefrac{3}{2}$ nature, the $\Delta$-baryon consists exclusively of an isospin-1 axialvector diquark .
              If the nucleon were made of a scalar diquark only, the $N\Delta\gamma$ transition would be a pure axial-scalar transition
              and the quark impulse-approximation in Fig.~(\ref{fig:current}a) would not participate.
              Table~\ref{tab:ax} shows that there are sizeable contributions coming from the interaction of the $\Delta$ with the
              axialvector component in the nucleon. These correlations appear in all diagrams of Fig.~\ref{fig:current}
              and contribute $\sim 40\%$ to $G_M^\star(0)$.
              An even more pronounced effect is again visible in $G_E^\star(0)$ whose axial-scalar contribution is negative throughout the $Q^2$ range,
              whereas its axial-axial component is positive. The combination yields a positive value for $G_E^\star(0)$ and thus a negative value for $R_{EM}(0)$.

               %!!!!!!!!!!!!!!!!!!!!!!!!!!!!!!!!!!!!!!!!!!!!!!!!!!!!!!!!!!!!!!!!!!!!!!!!!!!!!!!!!!!!!!!!!!!!!!!!!!!!!!!!!!!!!!!

                \begin{figure*}[t]
                           \begin{center}
                           \includegraphics[scale=0.25]{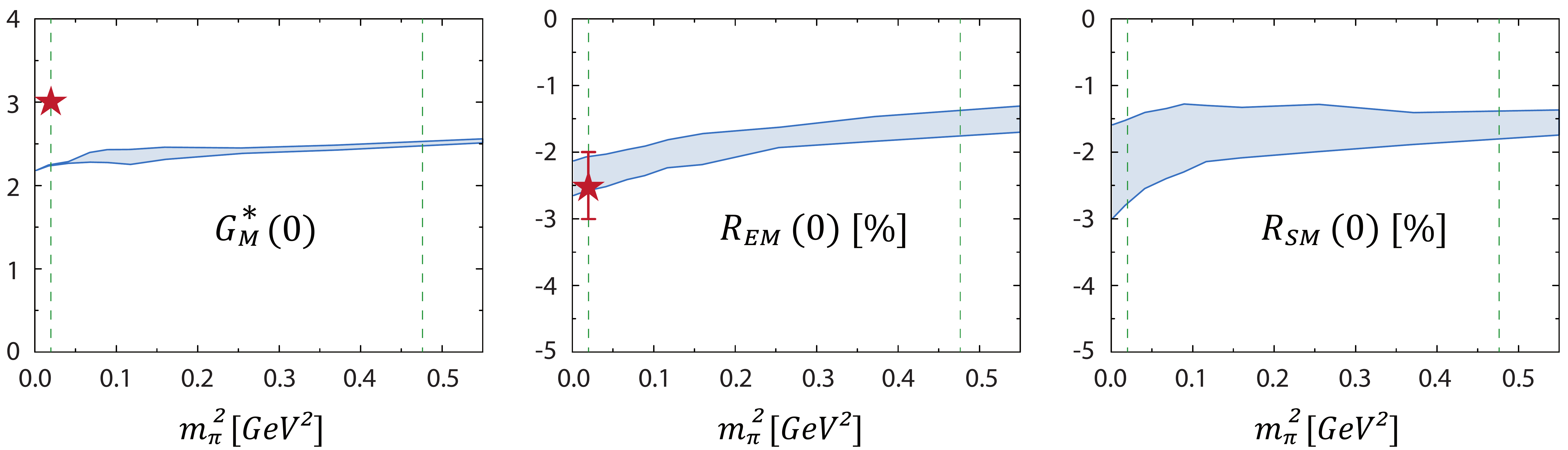}
                           \caption{(Color online)
                                    Current-quark mass dependence of the three form factors $G_M^\star$, $R_{EM}$ and $R_{SM}$ at vanishing momentum transfer.
                                    The dashed vertical lines indicate the position of the $u/d$ and strange-quark mass, and the stars are the experimental values from the PDG~\cite{Nakamura:2010zzi}.
                                    As there is no $s\bar{s}$ pseudoscalar meson in nature, the value $m_{s\conjg{s}}=0.69$ GeV corresponds to a meson-BSE solution
                                            at a strange-quark mass $m_s = 150$ MeV~\cite{Holl:2004fr}.            }\label{fig:mpi-dependence}
                           \end{center}
               \end{figure*}

               %!!!!!!!!!!!!!!!!!!!!!!!!!!!!!!!!!!!!!!!!!!!!!!!!!!!!!!!!!!!!!!!!!!!!!!!!!!!!!!!!!!!!!!!!!!!!!!!!!!!!!!!!!!!!!!!

                \renewcommand{\arraystretch}{1.0}

                \begin{table}[t]

                     \begin{center}

                     \begin{tabular}{    @{\;\;}c@{\;\;}  ||   @{\;\;}c@{\;\;} | @{\;\;}c@{\;\;} | @{\;\;}c@{\;\;}     }

                                                                 &  $G_M^\star(0)$  &  $G_E^\star(0)$  &  $G_C^\star(0)$       \\   \hline

                         $\Delta(\text{ax})-N(\text{sc})$        &  $1.27(4)$       &  $-0.05(2)$      &  $0.53(3)$                          \\
                         $\Delta(\text{ax})-N(\text{ax})$        &  $0.96(2)$       &  $0.11(1)$       &  $-0.11(3)$                 \\   \hline
                         Full                                    &  $2.23(2)$       &  $0.06(1)$       &  $0.42(6)$

                     \end{tabular}

                     \caption{Extrapolated results for the Jones-Scadron form factors at vanishing momentum transfer,
                              and their decomposition according to the diquark content of the nucleon's bound-state amplitude.
                              The first line corresponds to the scalar-diquark components in the nucleon and the second line to its axial-vector contributions.
                                            }\label{tab:ax}
                     \end{center}

                \end{table}

               %!!!!!!!!!!!!!!!!!!!!!!!!!!!!!!!!!!!!!!!!!!!!!!!!!!!!!!!!!!!!!!!!!!!!!!!!!!!!!!!!!!!!!!!!!!!!!!!!!!!!!!!!!!!!!!!

          \subsection{Quark-mass dependence}

              We now turn to the current-quark mass dependence of the $N\Delta\gamma$ transition form factors.
              It is accessible by modifying the current-quark mass that enters the quark DSE,
              and by implementing the corresponding quark propagator in all subsequent equations.
              The quark mass can be further mapped onto the pion mass by simultaneously solving the BSE for the pion.

              The results in Fig.~\ref{fig:mpi-dependence} show the extrapolated values of $G_M^\star$, $R_{EM}$ and $R_{SM}$
              at vanishing photon momentum. The error bands combine
              the accumulated uncertainties from the model parameter $\eta$, the seagull variation, and the extrapolation error.
              We find that, at least within the model uncertainties, the form factors $G_M^\star(0)$ and $R_{SM}(0)$
              are mostly insensitive to the pion-mass evolution.
              The only quantity that shows a more pronounced quark-mass dependence is the ratio $R_{EM}(0)$ which slowly decreases in magnitude.
              This might be again due to $p-$wave effects in the nucleon and $\Delta$ amplitudes which diminish with increasing quark masses
              and mainly have an impact on the electric quadrupole form factor.

              Further information on the pion-mass dependent properties of the form factors $G_M^\star$, $R_{EM}$ and $R_{SM}$
              is available from chiral effective field theory~\cite{Butler:1993ht,Pascalutsa:2005vq,Gail:2005gz}, see~\cite{Pascalutsa:2006up} for a review.
              Depending on the treatment of the small scales $m_\pi$ and $M_\Delta-M_N$,
              those results show different characteristics and typically exhibit a sizeable dependence on the pion mass.
              A consistent observation is that $R_{SM}$ logarithmically diverges in the chiral limit as a result of including pion loops.

              Since pion-cloud effects are missed in our approach, our results can provide insight in the properties of the quark-core contribution.
              Its insensitivity to a quark-mass variation, together with the overall agreement of $R_{EM}$ and $R_{SM}$ with experimental data at non-zero $Q^2$, suggests
              that the bulk contribution to these ratios is generated by the dynamics of the quark core.
              Thus, the impact of genuine pion-cloud effects in the form-factor ratios might be smaller than expected and mainly appear close to the chiral limit.
              Naturally, this question demands further investigation.

%%%%%%%%%%%%%%%%%%%%%%%%%%%%%%%%%%%%%%%%%%%%%%%%%%%%%%%%%%%%%%%%%%%%%%%%%%%%%%%%%%%%%%%%%%%%%%%%%%%%%%%%%%%%%%%%%%%%%%%%%%%%

\section{Conclusions}

              We investigated the $N \to \Delta \gamma$ transition properties in a Poincar\'e-covariant quark-diquark approach.
              Since pion-cloud effects are not yet incorporated, our results describe the quark core contributions to the transition form factors.
              They are obtained by employing a rainbow-ladder truncation at the level of the dressed quark propagator which amounts to a dressed-gluon exchange
              between the quarks inside the diquark. Thereafter, all quark and diquark ingredients are self-consistently determined from
              the corresponding Dyson-Schwinger and Bethe-Salpeter equations and thus related to the fundamental quantities in QCD.

              We find several characteristic features with impact for future investigations of the $N\to\Delta\gamma$ transition.
              As a consequence of Poincar\'e covariance, the electric quadrupole transition is dominated by $p-$wave admixtures
              to the nucleon and $\Delta$ bound-state amplitudes.
              Upon removing $p$ waves, the $E2/M1$ ratio becomes positive and grows toward the perturbative prediction $R_{EM}=1$.
              The magnetic dipole and Coulomb quadrupole transitions, on the other hand, are generated from $s-$wave contributions alone.
              The contribution from $d$ waves is almost negligible in all cases.

              All form factors depend only weakly on the current-quark mass.
              The dominant transition in the $N\to \Delta \gamma$ reaction, the magnetic dipole $M1$,
              could receive $\sim 25\%$ pion-cloud corrections in proximity to the chiral limit.
              On the other hand, the $E2/M1$ and $C2/M1$ ratios reproduce the experimental data quite well.
              These results disfavor arguments which promote the pion cloud and/or $d-$wave contributions as the prevailing missing effects
              for a proper description of the non-spherical character of the transition.
              Instead, this feature is already encoded in the nucleon and $\Delta$ quark core, partly via $s-$waves and partially via quark orbital angular momentum in terms of $p$ waves.

              From a future perspective, in alignment with the latest state-of-the-art calculations for nucleon and $\Delta$ masses and form factors,
              our approach can be improved by abandoning the diquark ansatz in favor of a Poincar\'e-covariant solution of the three-quark Faddeev equation.
              Apart from eliminating the model dependence on diquark properties, such an investigation would also be able to reach higher photon momenta than presently feasible.
              Furthermore, a consistently developed truncation beyond rainbow-ladder will methodically enable the inclusion of missing chiral corrections.
              This will elucidate the understanding of nucleon resonances and the interplay between pion-cloud contributions and the core of quarks.

% and Coulomb $C2$

%%%%%%%%%%%%%%%%%%%%%%%%%%%%%%%%%%%%%%%%%%%%%%%%%%%%%%%%%%%%%%%%%%%%%%%%%%%%%%%%%%%%%%%%%%%%%%%%%%%%%%%%%%%%%%%%%%%%%%%%%%%%

     \section{Acknowledgements}

     G.~E. is grateful to R. Alkofer, C. S. Fischer, and A. Krassnigg for valuable discussions.
     D.~Nicmorus thanks D. Rischke and J. Drobnak for their active support.
     This work was supported by the Austrian Science Fund FWF under Erwin-Schr\"odinger-Stipendium No.~J3039,
     and by the Helmholtz International Center for FAIR within the LOEWE program of the State of Hesse.
     G.~E. acknowledges support from FWF Project No.~P20592-N16.

%%%%%%%%%%%%%%%%%%%%%%%%%%%%%%%%%%%%%%%%%%%%%%%%%%%%%%%%%%%%%%%%%%%%%%%%%%%%%%%%%%%%%%%%%%%%%%%%%%%%%%%%%%%%%%%%%%%%%%%%%%%%%%%%%%%%%%%%%%%%%%%%%%%%%%%%%%%%%%%%%%%%%%%%%%%%%
%%%%%%%%%%%%%%%%%%%%%%%%%%%%%%%%%%%%%%%%%%%%%%%%%%%%%%%%%%%%%%%%%%%%%%%%%%%%%%%%%%%%%%%%%%%%%%%%%%%%%%%%%%%%%%%%%%%%%%%%%%%%%%%%%%%%%%%%%%%%%%%%%%%%%%%%%%%%%%%%%%%%%%%%%%%%%
%%%%%%%%%%%%%%%%%%%%%%%%%%%%%%%%%%%%%%%%%%%%%%%%%%%%%%%%%%%%%%%%%%%%%%%%%%%%%%%%%%%%%%%%%%%%%%%%%%%%%%%%%%%%%%%%%%%%%%%%%%%%%%%%%%%%%%%%%%%%%%%%%%%%%%%%%%%%%%%%%%%%%%%%%%%%%

\begin{appendix}

  \section{Euclidean conventions} \label{app:conventions}

            We work in Euclidean momentum space with the following conventions:
            \begin{equation}
                p\cdot q = \sum_{k=1}^4 p_k \, q_k,\quad
                p^2 = p\cdot p,\quad
                \Slash{p} = p\cdot\gamma\,.
            \end{equation}
            A vector $p$ is spacelike if $p^2 > 0$ and timelike if $p^2<0$.
            The hermitian $\gamma-$matrices $\gamma^\mu = (\gamma^\mu)^\dag$ satisfy the anticommutation relations
            $\left\{ \gamma^\mu, \gamma^\nu \right\} = 2\,\delta^{\,\mu\nu}$, and we define
            \begin{equation}
                \sigma^{\mu\nu} = -\frac{i}{2} \left[ \gamma^\mu, \gamma^\nu \right]\,, \quad
                \gamma^5 = -\gamma^1 \gamma^2 \gamma^3 \gamma^4\,.
            \end{equation}
            In the standard representation one has:
            \begin{equation*}
                \gamma^k  =  \left( \begin{array}{cc} 0 & -i \sigma^k \\ i \sigma^k & 0 \end{array} \right), \;
                \gamma^4  =  \left( \begin{array}{c@{\quad}c} \mathds{1} & 0 \\ 0 & \!\!-\mathds{1} \end{array} \right), \;
                \gamma^5  =  \left( \begin{array}{c@{\quad}c} 0 & \mathds{1} \\ \mathds{1} & 0 \end{array} \right),
            \end{equation*}
            where $\sigma^k$ are the three Pauli matrices.
            The charge conjugation matrix is given by
            \begin{equation}
                C = \gamma^4 \gamma^2, \quad C^T = C^\dag = C^{-1} = -C\,,
            \end{equation}
            and the charge conjugates for (pseudo-)\,scalar, \mbox{(axial-)} vector and tensor amplitudes are defined as
            \begin{equation}\label{chargeconjugation}
            \begin{split}
                \conjg{\Gamma}(p,P) &:= C\,\Gamma(-p,-P)^T\,C^T \,,   \\
                \conjg{\Gamma}^\alpha(p,P) &:= -C\,{\Gamma^\alpha}(-p,-P)^T\,C^T \,,   \\
                \conjg{\Gamma}^{\beta\alpha}(p,P) &:= C\,{\Gamma^{\alpha\beta}}(-p,-P)^T\,C^T\,,
            \end{split}
            \end{equation}
            where $T$ denotes a Dirac transpose.
            Four-momenta are conveniently expressed through hyperspherical coordinates:
            \begin{equation}\label{APP:momentum-coordinates}
                p^\mu = \sqrt{p^2} \left( \begin{array}{l} \sqrt{1-z^2}\,\sqrt{1-y^2}\,\sin{\phi} \\
                                                           \sqrt{1-z^2}\,\sqrt{1-y^2}\,\cos{\phi} \\
                                                           \sqrt{1-z^2}\;\;y \\
                                                           \;\; z
                                         \end{array}\right),
            \end{equation}
            and a four-momentum integration reads:
            \begin{equation*} \label{hypersphericalintegral}
                 \int\limits_p := \frac{1}{(2\pi)^4}\,\frac{1}{2}\int\limits_0^{\infty} dp^2 \,p^2 \int\limits_{-1}^1 dz\,\sqrt{1-z^2}  \int\limits_{-1}^1 dy \int\limits_0^{2\pi} d\phi \,.
            \end{equation*}

  \section{Partial-wave decomposition of nucleon and $\Delta$ quark-diquark amplitudes} \label{app:partial-wave}

            This appendix serves as a supplement to Section~\ref{sec:amplitudes}.
            We will highlight the general structure of the basis elements $\tau_k$, $\tau_k^\mu$ and $\tau_k^{\mu\rho}$
            that enter the quark-diquark amplitudes in Eqs.~\eqref{nuc:amplitudes} and~\eqref{delta:amplitudes}
            as well as their partial-wave decomposition.

            To count the number of basis elements in the nucleon and $\Delta$ bound-state amplitudes, we observe that
            $\Gamma_\text{N}^0$ and $\Gamma_\text{N}^\mu$ are three-point functions with two spinor legs (for quark and nucleon) and a scalar or axial-vector diquark leg.
            $\Gamma_\Delta^{\mu\nu}$ is a four-point function with two spinor indices, an axialvector-diquark index $\mu$ and a further Rarita-Schwinger index $\nu$.
            The largest linearly independent set of basis elements for each case reads:
            \begin{align}
            \Gamma_N^0: & \quad \{ \mathds{1}, \, \Slash{r}, \, \hat{\Slash{P}}, \, \Slash{r}  \, \hat{\Slash{P}} \}, \nonumber \\
            \Gamma_N^\mu: & \quad\{ \gamma_T^\mu, \,r^\mu, \,\hat{P}^\mu \} \times \{ \mathds{1}, \, \Slash{r}, \, \hat{\Slash{P}}, \, \Slash{r}  \, \hat{\Slash{P}} \}, \label{qdq-basis-general}\\
            \Gamma_\Delta^{\mu\nu}:  & \quad  \delta^{\mu\nu} \times\{ \mathds{1}, \, \Slash{r}, \, \hat{\Slash{P}}, \, \Slash{r}  \, \hat{\Slash{P}} \}, \nonumber \\
                                    & \quad\{ \gamma_T^\mu, \,r^\mu, \,\hat{P}^\mu \} \times \{ \gamma_T^\nu, \,r^\nu, \,\hat{P}^\nu \} \times\{ \mathds{1}, \, \Slash{r}, \, \hat{\Slash{P}}, \, \Slash{r}  \, \hat{\Slash{P}} \}. \nonumber
            \end{align}
            In case of the nucleon, the projector $\Lambda_+(P)$ in~\eqref{nuc:amplitudes} absorbs all occurrences of $\hat{\Slash{P}}$ via Eq.~\eqref{projector-contractions}
            so that two independent basis elements remain for the scalar quark-diquark amplitudes and six elements for the axial-vector contributions.
            The same applies to the $\Delta$, where the Rarita-Schwinger projector in addition
            eliminates all instances of $\gamma_T^\nu$ and $\hat{P}^\nu$ which leaves the eight basis elements in Eq.~\eqref{qdq-basis-onshell}.

            To obtain the decomposition in Table~\ref{table:basis} we need to construct quark-diquark spin and orbital angular-momentum operators.
            We will illustrate the procedure here for the case where only the dominant diquark amplitudes $\gamma^5 C$ and $\gamma^\mu C$ are retained.
            While the total angular momentum $j$ of the baryon, corresponding to the eigenvalues of the squared Pauli-Lubanski operator,
            is Poincar\'e-invariant, the classification in terms of quark-diquark spin and orbital angular momentum is not.
            As exemplified in Ref.~\cite{Eichmann:2011vu} in the three-body framework, one can retain a covariant notation by introducing formally covariant three-quark operators $S^\mu$ and $L^\mu$
            which in the rest frame coincide with the spin and orbital angular-momentum operators. The square of $S^\mu$ then reads
            \begin{equation}\label{operators-1}
              S^2  = \textstyle\frac{9}{4} \displaystyle \,\mathds{1} \otimes \mathds{1} \otimes \mathds{1} + \textstyle\frac{1}{4} \displaystyle \left( \sigma_T^{\mu\nu} \otimes \sigma_T^{\mu\nu} \otimes \mathds{1} + \text{perm.} \right)\,.
            \end{equation}

         Applying \eqref{operators-1} to the basic scalar and axialvector diquark structures $\gamma^5 C$ and $\gamma^\mu C$ yields
         \begin{align}
             S^2_{\alpha\alpha'\beta\beta'\gamma\gamma'} (\gamma^5 C)_{\beta'\gamma'} &=
                 \big[\textstyle\frac{3}{4}\,\mathds{1}\big]_{\alpha\alpha'}
                 (\gamma^5 C)_{\beta\gamma}\,, \label{S2-scalar}\\
             S^2_{\alpha\alpha'\beta\beta'\gamma\gamma'} (\gamma^\mu C)_{\beta'\gamma'} &= \nonumber\\
                =\Big[\textstyle\frac{3}{4}\,\delta^{\mu\nu} + 3 \,\Big( T_P^{\mu\nu}-&\textstyle\frac{1}{3}\,\gamma^\mu_T \,\gamma^\nu_T \Big)\Big]_{\alpha\alpha'}
                 (\gamma^\nu C)_{\beta\gamma}\,. \label{S2-axial}
         \end{align}
         and thereby defines effective quark-diquark spin operators,
         namely the square brackets in the above equation,
         that act on the quark index $\alpha$ and the axialvector diquark index $\mu$.
         They yield eigenvalues $s(s+1)$
         for appropriate linear combinations of \eqref{qdq-basis-onshell}.
         In the scalar-diquark case \eqref{S2-scalar}, only $s=\nicefrac{1}{2}$ can appear.
         In the axial-vector diquark case \eqref{S2-axial}, the spin operator is the sum
         of spin-$\nicefrac{1}{2}$ and spin-$\nicefrac{3}{2}$ projectors:
         \begin{equation}
             \big[\textstyle\frac{3}{4}\left(\delta^{\mu\nu}-\Lambda^{\mu\nu}\right) + \frac{15}{4} \,\Lambda^{\mu\nu}\big]_{\alpha\alpha'}\,,
         \end{equation}
         where $\Lambda^{\mu\nu}=T_P^{\mu\nu}-\textstyle\frac{1}{3}\gamma^\mu_T \gamma^\nu_T$.
         Thus, for instance, basis elements that exclusively depend on the quantities  $\hat{P}^\mu$ or $\gamma_T^\mu$
         must carry $s=\nicefrac{1}{2}$ as they are orthogonal to $\Lambda^{\mu\nu}$, cf. Eq.~\eqref{projector-contractions}.

         The quark-diquark orbital angular momentum is encoded in the operator
         \begin{equation}
             L^2= 2 p_T\cdot \partial_p + \left( p_T^\mu\,p_T^\nu - p_T^2\,T_P^{\mu\nu} \right) \partial^\mu_p\,\partial^\nu_p
         \end{equation}
         with eigenvalues $l(l+1)$. In the baryon's rest frame, it reduces to the usual orbital angular-momentum operator
         \begin{equation}
             L^2 =   2 \,\vect{p} \cdot\grad_{\vect{p}}  +p^k (\vect{p} \cdot\grad_{\vect{p}}) \,\nabla^k_{\vect{p}}  -\mathbf{p}^2 \Delta_{\vect{p}} \,.
         \end{equation}
         Only the action of $L^2$ on the basis elements $\tau_k(r,\hat{P})$ is relevant since the dressing functions of Eqs.~(\ref{nuc:amplitudes}--\ref{delta:amplitudes}) are Lorentz-invariant, and
         $L^2$ applied to any of the Lorentz-invariants $p^2$ and $z=\hat{p}\cdot\hat{P}$ yields zero.
           Moreover, the eigenstates of $L^2$ can be determined independently of their Dirac structure since $L^2$
           only acts upon the relative-momentum dependence.
           The basis elements in Eq.~\eqref{qdq-basis-onshell} involve up to three powers in the momenta $r^\alpha$.
           The corresponding eigenstates of $L^2$ are given by
           \begin{equation}\label{L2-eigenstates}
           \begin{split}
               l=1: & \quad r^\alpha \\
               l=2: & \quad 3\,r^\alpha r^\beta - T_P^{\alpha\beta} \\
               l=3: & \quad 5\,r^\alpha r^\beta r^\gamma - \mathcal{S}_{[\alpha\beta\gamma]} \big( T_P^{\alpha\beta}r^\gamma\big)
           \end{split}
           \end{equation}
           where $\mathcal{S}_{[\alpha\beta\gamma]}$ denotes a symmetric permutation of the indices $\alpha$, $\beta$, $\gamma$.

           Reexpressing the relative-momentum dependence of the basis elements in terms of \eqref{L2-eigenstates} allows to read off their orbital angular momentum content.
           In the case of the nucleon these structures are already visible in Eq.~\eqref{qdq-basis-onshell}:
           \begin{itemize}
           \item  $\mathds{1}$, $\gamma_T^\mu$ and $\hat{P}^\mu$ carry no relative-momentum dependence and correspond to $l=0$;
           \item  $\Slash{r}$, $r^\mu$, $\gamma_T^\mu\,\Slash{r}$ and $\hat{P}^\mu \Slash{r}$ are proportional to $r$ and therefore $p$ waves;
           \item and $r^\mu \Slash{r}$ must be replaced by $3 r^\mu \Slash{r} - \gamma_T^\mu$ to obtain a $d$ wave.
           \end{itemize}
           The procedure is analogous for the $\Delta$ baryon, where $\delta^{\mu\nu}$ is an $s$ wave,
           $\{ \gamma_T^\mu\,r^\nu,\, \hat{P}^\mu r^\nu, \, \delta^{\mu\nu} \Slash{r}\}$
           are $p$ waves and so on. Combined with the respective spin eigenstates, one arrives at the decomposition of Fig.~\ref{fig:partial-wave} and Table~\ref{table:basis}.
           The structure dictated by Eq.~\eqref{L2-eigenstates} is not immediately apparent in the $\Delta$ covariants since we have exploited the
           properties~\eqref{projector-contractions} of the Rarita-Schwinger projector that appears in the full amplitudes.

  \section{Electromagnetic transition current} \label{sec:current}

        \subsection{Kinematics}\label{app:kinematics}

             The $N\Delta\gamma$ transition matrix of Eq.~\eqref{current-general-0a} involves two independent momenta,
             the incoming nucleon momentum $P_i$ and outgoing $\Delta$ momentum $P_f$. They can be expressed by the photon momentum $Q$ and
             the average momentum $P$:
             \begin{equation}\label{current-kinematics}
                 Q=P_f-P_i\,, \quad
                 P = \frac{P_i + P_f}{2}\,,
             \end{equation}
             with the inversion $P_i = P-Q/2$ and $P_f = P+Q/2$.
             Only $Q^2$ remains as an independent variable
             since the nucleon and $\Delta$ are onshell: $P_i^2 = -M_N^2$ and $P_f^2=-M_\Delta^2$.
             For the remaining Lorentz-invariant combinations one obtains:
             \begin{equation*}
                 P^2 = -\frac{M_\Delta^2+M_N^2}{2}-\frac{Q^2}{4}\,, \quad
                 P\cdot Q = -\frac{M_\Delta^2-M_N^2}{2}\,,
             \end{equation*}
             and $P_i\cdot P_f = P^2-Q^2/4$.
             To simplify the notation, we abbreviate
             \begin{equation}\label{kinematics-variables}
                 M^2:= \frac{M_\Delta^2+M_N^2}{2} , \;\;
                 \delta := \frac{M_\Delta^2-M_N^2}{4M^2} , \;\;
                 \tau := \frac{Q^2}{4M^2}
             \end{equation}
             from which we obtain
             \begin{equation}\label{PdotQ}
             \begin{split}
                 P^2 &= -M^2 (1+\tau)\,, \\
                 P\cdot Q &= -2M^2 \delta\,, \\
                 P_i\cdot P_f &= -M^2 (1+2\tau)\,.
             \end{split}
             \end{equation}
             Expressed in terms of $\delta$ and $\tau$, the quantities defined in Eq.~\eqref{tau-lambda} are given by
             \begin{equation}\label{lambda-omega-simple}
                 \lambda_\pm = \frac{1\pm \sqrt{1-4\delta^2}}{2}+\tau  \,, \quad
                 \omega =\sqrt{\delta^2+\tau(1+\tau)} \,,
             \end{equation}
             and the normalized transverse average momentum $K$ of Eq.~\eqref{PT} becomes
             \begin{equation}\label{Ktransverse}
                 K^\mu = \widehat{P_T}^\mu = \frac{\sqrt{\tau}}{iM\omega}\left( P^\mu + \frac{\delta}{2\tau}\,Q^\mu\right).
             \end{equation}
             Vice versa, $P$ and $Q$ depend on $K$ and $\widehat{Q}$ via
             \begin{equation}\label{PQ-vs-KQHat}
                 P^\mu = \frac{M}{\sqrt{\tau}}\left( i\omega  K^\mu - \delta\, \widehat{Q}^\mu \right), \quad
                 Q^\mu = 2M\sqrt{\tau}\,\widehat{Q}^\mu\,.
             \end{equation}
             These expressions are ideally suited for evaluation in the the frame where $\widehat{Q} = e_3$ and $K = e_4$
             are simply the Euclidean unit four-vectors.

        \subsection{Structure of the current} \label{sec:currentgeneral}

             The generic structure of the matrix-valued $N\Delta\gamma$ transition current is given in Eq.~\eqref{current-general-0a}.
             It involves $\Gamma^{\mu,\alpha}(P,Q)$ which is a four-point function of positive parity, with two spinor and two vector indices ($\mu$ is the photon and $\alpha$ the
             Rarita-Schwinger index). Its most general Poincar\'e-covariant tensor structure involves 40 elements:
             \begin{equation}\label{current-basis-general}
             \left\{ \begin{array}{c} \gamma^\alpha \gamma^\mu \\ \delta^{\alpha\mu} \end{array}\;
                     \begin{array}{c} \gamma^\alpha P^\mu \\ \gamma^\alpha Q^\mu \\ P^\alpha \gamma^\mu  \\ Q^\alpha \gamma^\mu \\  \end{array}\;
                     \begin{array}{c} P^\alpha P^\mu \\ P^\alpha Q^\mu \\ Q^\alpha P^\mu \\ Q^\alpha Q^\mu \\  \end{array}
             \right\}
             \times  \left\{ \mathds{1}, \, \Slash{P},\, \Slash{Q},\, \left[\Slash{P},\,\Slash{Q}\right] \right\}.
             \end{equation}
             Only a few of those will survive on the $N$ and $\Delta$ mass shell which is enforced through the projectors in Eq.~\eqref{current-general-0a}.
             Using the properties
             \begin{equation}
             \begin{split}
                 &\Lambda_+(P_f) \,\Slash{P}_f = iM_\Delta\,\Lambda_+(P_f)\,,  \\
                 &\Slash{P}_i \,\Lambda_+(P_i) = iM_N\,\Lambda_+(P_i)
             \end{split}
             \end{equation}
             of Eq.~\eqref{projector-contractions} allows to systematically eliminate the structures $\Slash{P}$, $\Slash{Q}$ and $\left[\Slash{P},\,\Slash{Q}\right]$,
             so that in the second bracket  of Eq.~\eqref{current-basis-general} only the unit matrix remains as an independent element.
             Moreover, the properties of the Rarita-Schwinger projector,
             \begin{equation}\label{projector-contractions-2}
                 \mathds{P}^{\rho\alpha}(P_f)\,\gamma^\alpha = \mathds{P}^{\rho\alpha}(P_f)\,P_f^\alpha = 0\,,
             \end{equation}
             and thus
             \begin{equation}\label{RSproperties1}
                 \mathds{P}^{\rho\alpha}(P_f)\,P^\alpha = -\mathds{P}^{\rho\alpha}(P_f)\,\frac{Q^\alpha}{2}\,,
             \end{equation}
             remove all occurrences of $\gamma^\alpha$ or $P^\alpha$ in \eqref{current-basis-general} as well.
             One remains with the following four basis elements:
             \begin{equation}\label{current-basis-remaining}
                  Q^\alpha \gamma^\mu, \quad Q^\alpha P^\mu, \quad \delta^{\alpha\mu}, \quad Q^\alpha Q^\mu,
             \end{equation}
             each of which carries a Lorentz-invariant form factor that depends on $Q^2$.
             Current conservation $Q^\mu J^{\mu,\rho}=0$ leads to a linear relation between them and
             implies that $J^{\mu,\rho}$ can only involve basis elements that are transverse to $Q$ in the index $\mu$.
             The same effect can be achieved by
             applying a transverse projector with respect to the photon momentum to~\eqref{current-basis-remaining}.
             With an additional normalization of the momenta,
             this results in the expression given in Eq.~\eqref{current-general1}:
             \begin{equation}
                 \Gamma^{\alpha\mu} = i \widehat{Q}^\alpha \left( g_1 \gamma^\mu_T + g_2 \,K^\mu\right) - g_3\,T_Q^{\alpha\mu} \,.
             \end{equation}

             Next, we want to verify the equivalence of the two forms for the transition current obtained from Eq.~\eqref{current-general1} with Eq.~\eqref{current-general-0b}.
             To this end, we must show that the tensor structures that appear in~\eqref{current-general-0b} can be reexpressed by those in~\eqref{current-general1}
             when contracted with the Rarita-Schwinger and positive-energy projectors.
             For notational convenience we abbreviate $\mathds{P}^{\rho\alpha}_f = \mathds{P}^{\rho\alpha}(P_f)$ and $\Lambda_i=\Lambda_+(P_i)$
             in the following.

             The structure $\widehat{Q}^\alpha K^\mu$ attached to $G_C^\star$ requires no further examination
             as it appears in both equations. The relation
             \begin{equation}\label{K-RS}
                 \mathds{P}^{\rho\alpha}_f K^\alpha = -\frac{i}{\omega}\,(\delta-\tau)\, \mathds{P}^{\rho\alpha}_f \hat{Q}^\alpha
             \end{equation}
             that follows from \eqref{Ktransverse} and \eqref{RSproperties1} allows to relate the tensorial structure of $G_E^\star$,
             \begin{equation}
                 T_Q^{\alpha\gamma}\,T_K^{\gamma\mu} = T_Q^{\alpha\mu} - K^\alpha K^\mu\,,
             \end{equation}
             to $T_Q^{\alpha\mu}$ and $\widehat{Q}^\alpha K^\mu$ as well.
             The remaining term including the Levi-Civita symbol can be written as
             \begin{equation}
             \begin{split}
                 \varepsilon^{\alpha\mu\gamma\delta}  K^\gamma \hat{Q}^\delta  &= \gamma_5 \Big[ ( T_Q^{\alpha\mu}-\gamma^\alpha_T \gamma^\mu_T) \Slash{K} \\
                                      & \qquad\quad  + K^\mu \gamma_T^\alpha - K^\alpha \gamma_T^\mu\Big] \,\hat{\Slash{Q}}\,,
             \end{split}
             \end{equation}
             cf.~Eq.~(A14) of Ref.~\cite{Eichmann:2011vu}. The terms with $\gamma^\alpha_T$ amount to
             \begin{equation}
                 \mathds{P}^{\rho\alpha}_f \gamma_5 \gamma_T^\alpha =
                 -\mathds{P}^{\rho\alpha}_f \gamma_5\,\widehat{Q}^\alpha \widehat{\Slash{Q}}\,,
             \end{equation}
             via Eq.~\eqref{projector-contractions-2},
             and those featuring $K^\alpha$ can be again reduced using~\eqref{K-RS}.
             By expressing $K$ and $\widehat{Q}$ through $P_i$ and $P_f$ and using the relations~\eqref{projector-contractions} one further derives
             \begin{equation}
             \begin{split}
                 & \Lambda_f \gamma_5  \, \Slash{K} \,\hat{\Slash{Q}} \,\Lambda_i =   \frac{i\lambda_+}{\omega} \,\Lambda_f \gamma_5\,\Lambda_i\,, \\[2mm]
                 & \Lambda_f \gamma_5 \,\gamma^\mu_T \left[  \Slash{K} + \frac{i}{\omega}\,(\delta-\tau)\,   \hat{\Slash{Q}} \right] \Lambda_i = \\
                 & \qquad= \Lambda_f \gamma_5 \left[ \frac{\sqrt{\tau}\sqrt{1+2\delta}}{\omega}\,\gamma^\mu_T + 2 K^\mu\right]\Lambda_i \,.
             \end{split}
             \end{equation}
             Putting pieces together, one finally obtains:
             \begin{equation*}
             \begin{split}
                  &\mathds{P}^{\rho\alpha}_f   \varepsilon^{\alpha\mu\gamma\delta}  K^\gamma \hat{Q}^\delta\,\Lambda_i = \\
                  & = \mathds{P}^{\rho\alpha}_f \,\gamma_5 \left[ \frac{i\lambda_+}{\omega}\,T_Q^{\alpha\mu}
                    + \widehat{Q}^\alpha K^\mu + \frac{\sqrt{\tau}\,\sqrt{1+2\delta}}{\omega}\,\widehat{Q}^\alpha \gamma_T^\mu \right] \Lambda_i\,.
             \end{split}
             \end{equation*}
             This expression features the same tensor structures as Eq.~\eqref{current-general1}, and therefore
             the current constructed from Eq.~\eqref{current-general-0b} is indeed a reparametrization of Eq.~\eqref{current-general1}. % and the relations in Eq.~\eqref{a} follow.

             Taking into account the full set of terms in Eq.~\eqref{current-general-0b} finally allows to read off the relation between the
             $g_i(Q^2)$ from Eq.~\eqref{current-general1} and the Jones-Scadron form factors $G_M^\star$, $G_E^\star$ and $G_C^\star$:
             \begin{equation}\label{g-vs-JonesScadron}
             \begin{split}
                 g_1 &= b\, \frac{\sqrt{1+2\delta}\,\sqrt{\tau}}{2\lambda_+}\,(G_M^\star-G_E^\star)\,, \\
                 g_2 &= b \left[ \frac{\omega}{2\lambda_+}\,(G_M^\star-G_E^\star) - \frac{ \tau\,G_C^\star + (\delta-\tau)\,G_E^\star}{\omega}\right], \\
                 g_3 &= b\, \frac{G_M^\star+G_E^\star}{2}\,,
             \end{split}
             \end{equation}
             where $b = \sqrt{\tfrac{3}{2}}\,(1+M_\Delta/M_N) $.

        \subsection{Extraction of the form factors} \label{sec:ff-extraction}

             The final issue to address is how the form factors are extracted from the $N\Delta\gamma$ transition matrix~\eqref{current-general-0a}
             once $J^{\mu,\rho}$ has been computed from its underlying dynamics.
             The simplest Lorentz scalars that involve $J^{\mu,\rho}$ are obtained from the following Dirac traces and momentum contractions:
             \begin{equation}\label{current-traces}
             \begin{split}
                  s_1 &= \text{Tr}\left[ \gamma_5 \,J^{\mu,\rho} \,\gamma^\nu  \right] T_K^{\mu\nu} K^\rho   \,, \\[1mm]
                  s_2 &= \text{Tr}\left[ \gamma_5 \,J^{\mu,\rho}  \right] T_K^{\mu\rho}\,, \\[1mm]
                  s_3 &= \text{Tr}\left[ \gamma_5 \,J^{\mu,\rho}  \right] K^\mu  K^\rho \,,
             \end{split}
             \end{equation}
             and their relation with the Jones-Scadron form factors is:
             \begin{equation}
             \begin{split}
                 G_M^\star &= \frac{3\sqrt{1-4\delta^2}}{4i\,b\,\omega }\left[ \frac{\lambda_+}{\omega}\,s_2 + \frac{\sqrt{\tau}\,\sqrt{1+2\delta}}{\delta-\tau}\,s_1 \right], \\
                 G_E^\star &= \frac{\sqrt{1-4\delta^2}}{4\,b\,\omega}\left[ \frac{\lambda_+}{\omega}\,s_2 - \frac{\sqrt{\tau}\,\sqrt{1+2\delta}}{\delta-\tau}\,s_1 \right], \\
                 G_C^\star &= \frac{3\sqrt{1-4\delta^2}}{4i\,b\,\omega^2} \, \frac{\lambda_+\,(1+2\delta)}{\delta-\tau}\,s_3\,.
             \end{split}
             \end{equation}

             Eqs.~\eqref{current-traces} are most conveniently evaluated in the frame where $\hat{Q}^\mu = e_3^\mu$ and $K^\mu = e_4^\mu$.
             Here the $\mu=3$ component of $J^{\mu,\rho}$ is longitudinal to the photon momentum and vanishes because
             of current conservation (as long as the current is conserved microscopically).
             Therefore, the traces become
             \begin{equation}\label{current-traces-frame}
             \begin{split}
                 s_1 &=\text{Tr}\left[ \gamma_5 \, J^{i,4} \,\gamma^i \right], \\
                 s_2 &= \text{Tr}\left[ \gamma_5 \, J^{i,i}\right],  \\
                 s_3 &= \text{Tr}\left[ \gamma_5 \,J^{4,4}  \right],
             \end{split}
             \end{equation}
             where $i$ is summed over $i=1,2$.

    \section{Transition current in the quark-diquark framework} \label{app:current-details}

    \subsection{Current diagrams} \label{app:current-dirac}

            In the following we collect the ingredients of the $N\Delta\gamma$ transition current matrix
            in the quark-diquark model which are depicted in Fig.~\ref{fig:current}.
            The explicit form of the current is given by a sum of impulse-approximation diagrams (upper two panels in Fig.~\ref{fig:current})
            and two-loop contributions which represent the photon's coupling to the quark-diquark kernel (lower three panels):
            \begin{align}\label{ff:current2}
                J^{\mu,\rho}  &= \int    \conjg{\Gamma}_\Delta^{\rho\alpha}(p_f,P_f)\, (X_\text{q}+X_\text{dq})^{\mu,\alpha\beta}\, \Gamma_N^{\beta}(p_i,P_i) \,+ \nonumber \\
                                    & + \int\!\!\!\int  \conjg{\Gamma}_\Delta^{\rho\alpha}(p_f,P_f)\, X_\text{K}^{\mu,\alpha\beta}\, \Gamma_N^{\beta}(p_i,P_i) \,.
            \end{align}
            The quark-diquark amplitudes $\Gamma_\Delta^{\rho\alpha}$ and $\Gamma_N^{\beta}$
            are the solutions of the nucleon and $\Delta$ bound-state equations~\eqref{quark-diquark-bse}.
            $P_i$ and $P_f$ are incoming and outgoing on-shell momenta and $Q=P_f-P_i$ is the photon momentum, cf.~Eq.~\eqref{current-kinematics}.
            The relative momenta $p_i$ and $p_f$ are independent loop momenta in the two-loop diagrams;
            in the one-loop diagrams they are related to each other:
            $p_f-p_i=(1-\xi)\,Q$ for the quark diagram and $p_f-p_i=-\xi\,Q$ for the diquark diagram,
            where $\xi \in [0,1]$ is an arbitrary momentum-partitioning parameter which must be specified prior to solving the quark-diquark BSEs.
            $\alpha,\beta=1\dots 4$ are the Lorentz indices of the axialvector diquark in the $\Delta$ and nucleon amplitudes,
            and $\beta=0$ corresponds to the additional scalar-diquark component in the nucleon.

            The ingredients of Eq.\,\eqref{ff:current2} are given by
            \begin{align}\label{current-ingredients-1}
                X_\text{Q}^{\mu,\alpha\beta}  &=  S(p_+)\,\Gamma^\mu_\text{q}(\conjg{p},Q)\, S(p_-) \, D^{\alpha\beta}(\conjg{k}) \,,  \\
                X_\text{DQ}^{\mu,\alpha\beta} &=  S(\conjg{p}) \, D^{\alpha\alpha'}(k_+) \, \Gamma^{\mu,\alpha'\beta'}_{\text{D}\gamma}(\conjg{k},Q) \, D^{\beta'\beta}(k_-) \,, \nonumber \\
                X_\text{K}^{\mu,\alpha\beta}  &= D^{\alpha\alpha'}(k_+)\, S(p_+)\,K^{\mu,\alpha'\beta'} S(p_-) \,D^{\beta'\beta}(k_-)   \nonumber
            \end{align}
            and depend on the quark-photon vertex $\Gamma^\mu_\text{q}$ and the diquark-photon vertex $\Gamma^{\mu,\alpha\beta}_{\text{D}\gamma}$,
            where the latter includes an axial-axial contribution and an axial-scalar transition component for $\beta=0$.
            The quark and diquark momenta are:
            \begin{align*}
                p_- &= p_i+\xi\,P_i\,,    &   k_- &= -p_i + (1-\xi)\,P_i\;,  \\
                p_+ &= p_f+\xi\,P_f\,,    &   k_+ &= -p_f + (1-\xi)\,P_f\;,
            \end{align*}
            and we denote the average quark and diquark momenta that appear in the vertices by $\conjg{p}=(p_++p_-)/2$ and
            $\conjg{k}=(k_++k_-)/2$.

            The 'gauged' kernel $K^{\mu,\alpha\beta}$ contains the exchange-quark diagram and the seagull vertices $M^{\mu,\alpha}$:
            \begin{equation}
                K^{\mu,\alpha\beta} = \left( K_\text{EX} + K_\text{SG} + K_{\overline{\text{SG}}} \right)^{\mu,\alpha\beta}\;,
            \end{equation}
            with
            \begin{align}\label{current-ingredients-2}
                K_\text{EX}^{\mu,\alpha\beta}              \!&= \Gamma^\beta(r_+,k_-)\Big[ S(q_+) \,\Gamma^\mu_\text{q}(\conjg{q},Q)\, S(q_-) \Big]^T \conjg{\Gamma}^\alpha(r_-,k_+)   \nonumber \\
                K_\text{SG}^{\mu,\alpha\beta}              \!&= M^{\mu,\beta}(r'_+,k_-,Q)\, S(q_+)^T\,\conjg{\Gamma}^\alpha(r_-,k_+)\,, \nonumber\\
                K_{\overline{\text{SG}}}^{\mu,\alpha\beta} \!&= \Gamma^\beta(r_+,k_-)\,S(q_-)^T\,\conjg{M}^{\mu,\alpha}(r'_-,k_+,Q)\,,
            \end{align}
            and momenta:
            \begin{equation*}
                q_\pm = k_\pm-p_\mp\,, \quad
                r_\pm = \frac{p_\pm-q_\mp}{2}\,,\quad
                r'_\pm = \frac{p_\pm-q_\pm}{2}\,.
            \end{equation*}
            The color-flavor traces of the various contributions in Eqs.~(\ref{current-ingredients-1}--\ref{current-ingredients-2}) are detailed in App.~\ref{app:color-flavor}.

    \subsection{Kinematic restrictions} \label{app:kinematic-restrictions}

            The integrand of the $N\Delta\gamma$ transition matrix~\eqref{ff:current2} contains
            dressed quark and diquark propagators which are computed selfconsistently from Eqs.~\eqref{quarkdse} and~\eqref{diquark-propagator}.
            In rainbow-ladder truncation, the quark propagator exhibits singularities in the timelike complex plane
            and the diquark propagators have poles on the timelike real axis. Without taking residues into account explicitly,
            that singularity structure restricts the accessible momentum phase space in the integrals~\eqref{ff:current2}
            and leads to kinematic limitations in the  $Q^2$ range of the resulting form factors.

            The problem can be accessed by separating the real and imaginary parts of the quark and diquark momenta
            that enter the diagrams.
            As we have noted in the discussion of Eqs.~\eqref{PQ-vs-KQHat} and~\eqref{current-traces-frame}, it is advantageous to work in the frame where
            the normalized momenta $\widehat{Q}$ and $K$ that describe the current are the Euclidean unit vectors $\widehat{Q}=e_3$ and $K=e_4$.
            For spacelike values of $Q^2$, the photon momentum $Q^\mu$ is then purely real and the transverse average momentum $P_T$ is imaginary.
            This frame also suggests itself via the impulse-approximation relations $p_f-p_i \sim Q$ from the previous subsection:
            if the photon momentum is real, both relative momenta $p_i$ and $p_f$ can be chosen real as well (the loop momentum is always real)
            and it is sufficient to determine the $N$ and $\Delta$ amplitudes for $p_i^2$, $p_f^2 \in \mathds{R}_+$.

            The quark and diquark momenta depend on the momentum partitioning parameter $\xi \in [0,1]$ which can be used for maximizing the available momentum phase space.
            We exemplify the procedure for the quark momentum $p_-=p_i+\xi\,P_i$. In the frame where the photon momentum is real: $p_i \in \mathds{R}^4$, and
            the only imaginary contribution in $P_i = P-Q/2$ comes from the transverse component $P_T$, cf.~Eq.~\eqref{PQ-vs-KQHat}. Thus, $p_-$ can be written as
            \begin{equation}
                p^\mu_- = R^\mu + \xi\,P_T^\mu = R^\mu + \frac{M\xi\omega}{\sqrt{\tau}}\,i K^\mu\,,
            \end{equation}
            where $R^\mu$ is purely real and irrelevant for the further discussion.
            $p_-^2$ describes the interior of a parabola in the complex plane,
            \begin{equation}
                \left( t \pm i\,\frac{M\xi\omega}{\sqrt{\tau}}\right)^2\,, \quad t \in \mathds{R}_+\,,
            \end{equation}
            that intersects the real axis at $(p_-^2)_\text{max} = -(M \xi \omega)^2/\tau$.
            $M$ is here the average $N$-$\Delta$ mass defined in Eq.~\eqref{kinematics-variables}.
            On the other hand,
            the nearest singularities in the quark propagator (whether timelike-real or in the timelike complex plane)
            can be used to construct another parabola
            \begin{equation}
               \left( t \pm i m_q\right)^2\,, \quad t \in \mathds{R}_+\,,
            \end{equation}
            which defines the quark 'pole mass' $m_q$.
            Hence, in order to sample the quark propagator only in the kinematically safe region, one has to obey the condition
            \begin{equation}
                -\frac{(M \xi \omega)^2}{\tau} > -m_q^2 \quad \Rightarrow \quad \frac{\delta^2}{\tau} + \tau < \left( \frac{m_q}{\xi M}\right)^2 -1\,,
            \end{equation}
            where we have used Eq.~\eqref{lambda-omega-simple} in the second step.
            Because of the non-vanishing $N$-$\Delta$ mass difference, expressed by $\delta \neq 0$, this puts constraints on $\tau=Q^2/(4M^2)$
            from both above \textit{and} below.

            The analysis can be repeated for any quark or diquark momentum that enters the integrands~\eqref{current-ingredients-1} and
            yields the combined relation
            \begin{equation}\label{kinematic-restriction-1}
                \frac{\delta^2}{\tau} + \tau < h(\xi)^2 -1\,,
            \end{equation}
            where
            \begin{equation}\label{kinematic-restriction-h}
                h(\xi) = \text{min} \left[ \frac{m_q}{\xi M} \,, \; \frac{m_q}{|1-2\xi| M} \,, \;\frac{m_d}{(1-\xi) M}  \right],
            \end{equation}
            with $m_d = \text{min}(m_\text{sc},m_\text{ax},2m_q)$.
            In view of maximizing the kinematical region in Eq.~\eqref{kinematic-restriction-1},
            we are interested in the value $\xi_0$ that maximizes $h(\xi)$.
            For $m_q < m_d < 2m_q$, that value is given by
            \begin{equation}
                \xi_0 = (1+m_d/m_q)^{-1} \in [ 1/3 \dots 1/2],
            \end{equation}
            which leads to $h_0 = h(\xi_0) = (m_q+m_d)/M$. Via Eq.~\eqref{kinematic-restriction-1}, the resulting lower and upper limits for $Q^2$ become
            \begin{equation}
                \tau_\pm = \frac{ h_0^2-1}{2}\left[ 1 \pm \sqrt{1-\left(\frac{2\delta}{h_0^2-1}\right)^2}\right].
            \end{equation}
            Inserting the numerical values for the masses $m_q$, $m_\text{sc}$, $m_\text{ax}$, $M_N$ and $M_\Delta$
            yields a kinematically safe region that is centered around $Q^2 \sim 1 \dots 1.3$ GeV$^2$, depending on the model parameter $\eta$ in the effective coupling.

            We conclude with two remarks. First, while it would be possible to choose two different momentum partitioning parameters $\xi_N$ and $\xi_\Delta$
            in the calculation of the nucleon and $\Delta$ amplitudes and form factors, doing so would not relax the constraint in Eq.~\eqref{kinematic-restriction-h}: the optimal choice corresponds to $\xi_N=\xi_\Delta$.
            The second remark concerns the choice of reference frame.
            The current diagrams could be equally evaluated in the $\Delta$ rest frame where $\widehat{P_f}=e_4$ and where
            $p_f^2$ can be chosen as the real loop momentum.
            In that case, one has access to the limit $Q^2=0$ and also to the timelike region whereas the largest accessible spacelike value is smaller, $Q^2 \sim 1$~GeV$^2$.
            On the other hand, $Q$ and $p_i$ will become complex which necessitates a complex continuation of the nucleon bound-state amplitude.
            Such a procedure has been applied to compute the $N\Delta\pi$ transition in the current framework~\cite{Mader:2011zf}.

    \subsection{Color and flavor traces} \label{app:color-flavor}

        In this appendix we collect the color and flavor traces that appear in the quark DSE,
        the various bound-state equations and the $N\Delta\gamma$ transition matrix element.

        Each quark-gluon vertex in the kernel~\eqref{RLkernel} is equipped with a color factor $\lambda_i/2$, where the $\lambda_i$ are the eight Gell-Mann matrices.
        The color factors for the diquark amplitudes are given by $\varepsilon_{ABC}/\sqrt{6}$
        and those for the $N$ and $\Delta$ quark-diquark amplitudes by $\delta_{AC}/\sqrt{3}$,
        where $A,B$ are quark indices and $C$ is the diquark index.
        Tracing the color structure leads to a prefactor $\nicefrac{4}{3}$ in front of the integral that appears in
        the quark DSE~\eqref{quarkdse}, a factor $-\nicefrac{4}{3}$ for those in the meson BSE and the inhomogeneous BSE~\eqref{qpv-bse} for the quark-photon vertex,
        and a prefactor $\nicefrac{2}{3}$ in the diquark BSE~\eqref{diquark-bse}.
        The color traces in the form-factor diagrams of Fig.~\ref{fig:current} yield $+1$ for the impulse-approximation diagrams and $-1$ for the
        exchange and seagull diagrams.

        Considering the flavor case, we work in the $SU(2)$ isospin-symmetric limit with two
        degenerate quark flavors  $\mathsf{u}=\bigl( \begin{smallmatrix}1\\0 \end{smallmatrix}\bigr)$ and
        $\mathsf{d}=\bigl( \begin{smallmatrix}0\\1 \end{smallmatrix}\bigr)$.
        Out of those one can construct the following diquark flavor matrices:
        \begin{equation}\label{flavor:diquarks}
          \mathsf{s} = \left(\,  \frac{i\sigma_2}{\sqrt{2}} \;\bigg| \,
                                 \frac{\one + \sigma_3 }{2}\,,\,
                                 \frac{\sigma_1}{\sqrt{2}}\,,\,
                                 \frac{\one - \sigma_3}{2}\, \right)
        \end{equation}
        where $\mathsf{s}_0$ corresponds to the isoscalar (scalar) diquark and
        the three $\mathsf{s}_i$, with $i=1,2,3$, to the isotriplet (axialvector) diquark.
        The $\sigma_i$ are the Pauli matrices, and the flavor
        matrices are normalized to $\text{Tr}\{\mathsf{s}_i^\dag\,\mathsf{s}_j \}=\delta_{ij}$.

        The flavor factors for baryons in the quark-diquark picture are given by the
        Clebsch-Gordan coefficients according to the respective diquark content of the
        baryon. For proton and neutron one obtains
        \begin{equation} \label{flavor:nucleon}
        \begin{split}
         \mathsf{p} &= \left(\,\mathsf{u} \;\Big| \,\sqrt{\tfrac{ 2}{3}} \,\mathsf{d}\,,\, -\sqrt{\tfrac{1}{3}}\, \mathsf{u}\,,\,0\, \right),\\
         \mathsf{n} &= \left(\,\mathsf{d} \;\Big| \,0\,,\,\sqrt{\tfrac{ 1}{3}} \,\mathsf{d}\,,\, -\sqrt{\tfrac{2}{3}}\, \mathsf{u}\, \right) ,
         \end{split}
        \end{equation}
        where the first entry represents the isoscalar diquark and the remaining
        three the contributions from the isovector channel, in the same order as Eq.~\eqref{flavor:diquarks}.
        For example, the two mixed-antisymmetric and mixed-symmetric flavor tensors of the proton's three-quark amplitude are
        $\mathsf{s}_0 \otimes \mathsf{p}_0$ and $\sum_i \mathsf{s}_i \otimes \mathsf{p}_i$, respectively.
        On the other hand, the $\Delta$-baryons do not contain any contribution from the scalar diquark, and
        the corresponding Clebsch-Gordan construction for the $\Delta^+$ and the $\Delta^0$ (only these two are relevant for the $N\Delta\gamma$ transition) yields
        \begin{equation} \label{flavor:delta}
        \begin{split}
         &\Delta^{+}= \left(\,\sqrt{\tfrac{ 1}{3}} \,\mathsf{d}\,,\,\sqrt{\tfrac{2}{3}}\, \mathsf{u}\,,\,0\, \right), \\
         &\Delta^{0} = \left(\,0\,,\,\sqrt{\tfrac{2}{3}} \,\mathsf{d}\,,\,\sqrt{\tfrac{1}{3}}\, \mathsf{u}\, \right) .
        \end{split}
        \end{equation}

        The flavor traces in the nucleon and $\Delta$ quark-diquark BSEs~\eqref{quark-diquark-bse}
        depend on the type of involved diquarks. This leads to combined color-flavor factors $c^{(00)}  =-c^{(\alpha\beta)}=-\nicefrac{1}{2}$ and
        $c^{(0\alpha)}=c^{(\alpha 0)} = \nicefrac{\sqrt{3}}{2}$ for the nucleon,  whereas in the case of the $\Delta$ the color-flavor trace is given by $c^{(\alpha\beta)}=-1$.

        The flavor factors for the form-factor diagrams of Fig.~\ref{fig:current} are obtained by combining Eqs.~(\ref{flavor:diquarks}--\ref{flavor:delta})
        with the quark charge matrix $\mathsf{Q} = diag(q_u,q_d)$ which is attached to the quark-photon vertex.
        In the case of the $p\gamma\to\Delta^{+}$ transition, the three contributions in Fig.~(\ref{fig:current}a--c)
        yield the following traces:
        \begin{equation}
        \begin{split}
            \sum_i (\Delta^{+}_i)^\dag \,\mathsf{Q}  \,\mathsf{p}_i &\,, \quad
            \sum_{ij} \left[ (\Delta^{+}_i)^\dag \,\mathsf{p}_j \right] 2\,\text{Tr}\{ \mathsf{s}_i^\dag\,\mathsf{s}_j\,\mathsf{Q}^\dag \}, \\
            &\sum_{ij} (\Delta^{+}_i)^\dag \, \mathsf{s}_j\,\mathsf{Q}^\dag\,\mathsf{s}_i^\dag\,\mathsf{p}_j  \,.
        \end{split}
        \end{equation}
        Only the axial-vector diquark contributes to the $\Delta$ (hence $i=1,2,3$)
        whereas the nucleon has both scalar ($j=0$) and axial-vector diquark components ($j=1,2,3$).

        The charge-flavor structure of the seagull diagrams in Fig.~(\ref{fig:current}d--e)
        appears in combination with their Dirac components via Eq.~\eqref{seagulls-generic}.
        The flavor factors read:
        \begin{equation}
        \begin{split}
            e_- &=  \sum_{ij} (\Delta^{+}_i)^\dag \, \mathsf{s}_j\,\mathsf{Q}^\dag\,\mathsf{s}_i^\dag\,\mathsf{p}_j\,, \\
            e_+ &=  \sum_{ij} (\Delta^{+}_i)^\dag \, \mathsf{Q} \,\mathsf{s}_j \,\mathsf{s}_i^\dag\,\mathsf{p}_j\,, \\
            e_{dq} &=  \sum_{ij} \left[ (\Delta^{+}_i)^\dag \,\mathsf{s}_j \,\mathsf{s}_i^\dag\,\mathsf{p}_j \right] 2\,\text{Tr}\{ \mathsf{s}_j^\dag\,\mathsf{s}_j\,\mathsf{Q}^\dag \}
        \end{split}
        \end{equation}
        for the seagull vertices in Fig.~(\ref{fig:current}d), and
        \begin{equation}
        \begin{split}
            \conjg{e}_+ &=  \sum_{ij} (\Delta^{+}_i)^\dag \, \mathsf{s}_j\,\mathsf{s}_i^\dag\,\mathsf{Q} \,\mathsf{p}_j\,, \\
            \conjg{e}_- &=  \sum_{ij} (\Delta^{+}_i)^\dag \, \mathsf{s}_j \,\mathsf{Q}^\dag \,\mathsf{s}_i^\dag\,\mathsf{p}_j\,, \\
            \conjg{e}_{dq} &=  \sum_{ij} \left[ (\Delta^{+}_i)^\dag \,\mathsf{s}_j \,\mathsf{s}_i^\dag\,\mathsf{p}_j \right] 2\,\text{Tr}\{ \mathsf{s}_i^\dag\,\mathsf{s}_i\,\mathsf{Q}^\dag \}
        \end{split}
        \end{equation}
        for the conjugated seagulls that appear in Fig.~(\ref{fig:current}e).

        In combination with the color traces $+1$ for the impulse-approximation diagrams and $-1$ for the
        exchange diagrams, the final result for the $p\gamma\to\Delta^{+}$ transition matrix element becomes
        \begin{equation*} \label{flavor-final}
        \Lambda = -\textstyle{\frac{\sqrt{2}}{3}}\left[ J_\text{Q} + J_\text{DQ}  + J_\text{EX} + J_\text{SG} + J_{\conjg{\text{SG}}} \right],
        \end{equation*}
        where the individual contributions are given by
        \begin{align}
            J_\text{Q}  &= \Lambda_\text{Q}^\text{AA}\,, \nonumber \\
            J_\text{DQ} &= - \Lambda_\text{DQ}^\text{AA} +  \sqrt{3}\,\Lambda_\text{DQ}^\text{AS}\,,  \\
            J_\text{EX} &= \textstyle{\frac{1}{2}}\,\Lambda_\text{EX}^\text{AA} - \textstyle{\frac{\sqrt{3}}{2}}\,\Lambda_\text{EX}^\text{AS}\,, \nonumber \\
            J_\text{SG} &= \textstyle{\frac{1}{2}} (\Lambda_1^\text{AA} + \Lambda_2^\text{AA}) - \Lambda_3^\text{AA}  -\textstyle{\frac{\sqrt{3}}{2}}(\Lambda_1^\text{AS} - \Lambda_2^\text{AS})\,,  \nonumber \\
            J_{\conjg{\text{SG}}} &=   \conjg{\Lambda}_1^\text{AA}-\textstyle{\frac{1}{2}} (\conjg{\Lambda}_2^\text{AA} + \conjg{\Lambda}_3^\text{AA})   -\textstyle{\frac{\sqrt{3}}{2}}(\conjg{\Lambda}_2^\text{AS} - \conjg{\Lambda}_3^\text{AS})\,.  \nonumber
        \end{align}
        In this notation, the $\Lambda_i$ denote the Dirac parts obtained from Eqs.~\eqref{current-ingredients-1} and~\eqref{current-ingredients-2}, and
        the superscripts S and A refer to the scalar or axial-vector
        diquark content in the outgoing $\Delta$ (left) and incoming nucleon (right) amplitudes.
        The labels $\{1,2,3\}$ for the seagull contributions correspond to those in Eq.~\eqref{seagulls-generic}.
        Identical traces follow for the process $n\gamma\to \Delta^0$.

\end{appendix}

%%%%%%%%%%%%%%%%%%%%%%%%%%%%%%%%%%%%%%%%%%%%%%%%%%%%%%%%%%%%%%%%%%%%%%%%%%%%%%%%%%%%%%%%%%%%%%%%%%%%%%%%%%%%%%%%%%%%%%%%%%%%%%%%%%%%%%%%%%%%%%%%%%%%%%%%%%%%%%%%%%%%%%%%%%%%%
%%%%%%%%%%%%%%%%%%%%%%%%%%%%%%%%%%%%%%%%%%%%%%%%%%%%%%%%%%%%%%%%%%%%%%%%%%%%%%%%%%%%%%%%%%%%%%%%%%%%%%%%%%%%%%%%%%%%%%%%%%%%%%%%%%%%%%%%%%%%%%%%%%%%%%%%%%%%%%%%%%%%%%%%%%%%%
%%%%%%%%%%%%%%%%%%%%%%%%%%%%%%%%%%%%%%%%%%%%%%%%%%%%%%%%%%%%%%%%%%%%%%%%%%%%%%%%%%%%%%%%%%%%%%%%%%%%%%%%%%%%%%%%%%%%%%%%%%%%%%%%%%%%%%%%%%%%%%%%%%%%%%%%%%%%%%%%%%%%%%%%%%%%%

\bibliographystyle{apsrev4-1-mod}

\bibliography{lit-ndeltagamma-2}

\end{document}